\documentclass[11pt]{article}
\usepackage{axodraw}
\usepackage{epsfig}
\usepackage{amsfonts}
\usepackage{amsmath}
\usepackage{bbm}
\allowdisplaybreaks[1]

\input pix.sty

\newcommand{\aL}{a^{ }_\rmii{L}}
\newcommand{\aR}{a^{ }_\rmii{R}}
\renewcommand{\eq}{eq.~}
\renewcommand{\eqs}{eqs.~}
\renewcommand{\se}{sec.~}

\newcommand{\Nc}{N_{\rm c}}

\newcommand{\gammaE}{\gamma_\rmii{E}}

\newcommand{\rmO}{{\mathcal{O}}}
\newcommand{\bmu}{\bar\mu}

\def\lsi{\raise0.3ex\hbox{$<$\kern-0.75em\raise-1.1ex\hbox{$\sim$}}}
\def\gsi{\raise0.3ex\hbox{$>$\kern-0.75em\raise-1.1ex\hbox{$\sim$}}}
\newcommand{\lsim}{\mathop{\lsi}}
\newcommand{\gsim}{\mathop{\gsi}}

\newcommand{\sign}{\mathop{\mbox{sign}}}
\newcommand{\nF}{n_\rmii{F}}
\newcommand{\nB}{n_\rmii{B}}
 \renewcommand{\nF}[1]{n_\rmii{F{#1}}}
 \renewcommand{\nB}[1]{n_\rmii{B{#1}}}
\newcommand{\ga}{\gamma}
\newcommand{\rmii}[1]{{\mbox{\tiny\rm{#1}}}}
\newcommand{\rmiii}[1]{{\mbox{\tiny{$\scriptstyle{\rm#1}$}}}}
\newcommand{\re}{\mathop{\mbox{Re}}}
\newcommand{\im}{\mathop{\mbox{Im}}}

\newcommand{\Tint}[1]{{\hbox{$\sum$}\!\!\!\!\!\!\!\int\,}_{\!\!\!\!\raise-0.9ex\hbox{$\scriptstyle{#1}$}}}
\newcommand{\Tinti}[1]{{{\Sigma}\!\!\!\!\raise0.3ex\hbox{$\int$}_\rmii{${#1}$}}}

\newcommand{\bi}{\begin{itemize}}
\newcommand{\ei}{\end{itemize}}
\newcommand{\hide}[1]{ }
\newcommand{\bsl}[1]{\,\slash\!\!\!\!{#1}\,}

 \newcommand{\mufac}{ } %{\mu^{-2\epsilon}}
 \newcommand{\error}{ } 
\def\TAsc(#1,#2)(#3,#4,#5)%
{\SetWidth{2.0}\CArc(#1,#2)(#3,#4,#5)\SetWidth{1.0}}
\def\Lwidth{3}

\def\TAgl(#1,#2)(#3,#4,#5){\SetWidth{2.0}\PhotonArc(#1,#2)(#3,#4,#5){\Lwidth}%
{6.283 #3 mul 360 div #4 #5 sub #4 #5 sub mul sqrt mul Tdensity mul}%
\SetWidth{1.0}}
\def\TLgl(#1,#2)(#3,#4){\SetWidth{2.0}\Photon(#1,#2)(#3,#4){\Lwidth}
{#1 #3 sub #1 #3 sub mul #2 #4 sub #2 #4 sub mul add sqrt Tdensity mul}%
\SetWidth{1.0}}
\renewcommand{\picc}[1]{\;\parbox[c]{60pt}{\begin{picture}(60,30)(0,0)
\SetWidth{1.0}\SetScale{1.3} #1 \end{picture}}\; }
\def\Lwidth{1.3}

\newcommand{\picu}[1]{\;\parbox[c]{40pt}{\begin{picture}(50,30)(0,0)
\SetWidth{1.0}\SetScale{0.8} #1 \end{picture}}\; }
\def\EleB{\picu{%
 \Asc(30,5)(22.3,27,153)%
 \Aqu(30,25)(22.3,207,333)%
 \COval(10,15)(2,2)(0){Black}{Black}%
 \COval(50,15)(2,2)(0){Black}{Black}%
 \Asc(30,35.0)(8,0,360)%
}}
\def\EleC{\picu{%
 \Asc(30,5)(22.3,27,67)%
 \Asc(30,5)(22.3,113,153)%
 \Aqu(30,25)(22.3,207,333)%
 \COval(10,15)(2,2)(0){Black}{Black}%
 \COval(50,15)(2,2)(0){Black}{Black}%
 \CArc(30,26.3)(9,-90,270)%
 \CArc(30,26.3)(7,-90,270)%
}}
\def\EleD{\picu{%
 \Asc(30,5)(22.3,27,153)%
 \Aqu(30,25)(22.3,207,333)%
 \COval(10,15)(2,2)(0){Black}{Black}%
 \COval(50,15)(2,2)(0){Black}{Black}%
 \Agl(30,35.0)(8,0,360)%
}}
\def\EleE{\picu{%
 \Asc(30,5)(22.3,27,153)%
 \Aqu(30,25)(22.3,207,333)%
 \COval(10,15)(2,2)(0){Black}{Black}%
 \COval(50,15)(2,2)(0){Black}{Black}%
 \Agl(30,26.3)(9,190,350)%
}}
\def\EleF{\picu{%
 \Asc(30,5)(22.3,27,153)%
 \Aqu(30,25)(22.3,207,333)%
 \COval(10,15)(2,2)(0){Black}{Black}%
 \COval(50,15)(2,2)(0){Black}{Black}%
 \Agl(30,3.7)(9,10,170)%
}}
\def\EleG{\picu{%
 \Asc(30,5)(22.3,27,153)%
 \Aqu(30,25)(22.3,207,270)%
 \Aqu(30,25)(22.3,270,333)%
 \COval(10,15)(2,2)(0){Black}{Black}%
 \COval(50,15)(2,2)(0){Black}{Black}%
 \Lgl(30,2.7)(30,27.3)%
}}
\makeatletter \@addtoreset{equation}{section} \makeatother
\renewcommand{\theequation}{\arabic{section}.\arabic{equation}}
\makeatletter
\renewcommand\section{\@startsection {section}{1}{\z@}%
                                   {-5.5ex \@plus -1ex \@minus -.2ex}% bfr-
                                   {2.3ex \@plus.2ex}%
                                   {\normalfont\large\bfseries}}
\renewcommand\subsection{\@startsection{subsection}{2}{\z@}%
                                     {-3.25ex\@plus -1ex \@minus -.2ex}%
                                     {1.5ex \@plus .2ex}%
                                     {\normalfont\normalsize\bfseries}}
\renewcommand\thesection {\@arabic\c@section}
\renewcommand\thesubsection   {\thesection.\@arabic\c@subsection}
\renewcommand{\@seccntformat}[1]{%
\csname the#1\endcsname.\hspace{1.0em}}
\makeatother
\begin{document}

\flushbottom

\begin{titlepage}

\begin{flushright}
% DRAFT \\ 
BI-TP 2011/44\\
% arXiv:1008.3263\\ 
\vspace*{1cm}
\end{flushright}
\begin{centering}
\vfill

{\Large{\bf
 Thermal right-handed neutrino production rate \\[2mm] 
 in the non-relativistic regime
}} 

\vspace{0.8cm}

M.~Laine, %$^{\rm a}$, %%\footnote{laine@physik.uni-bielefeld.de}
Y.~Schr\"oder %%\footnote{yorks@physik.uni-bielefeld.de}

\vspace{0.8cm}

%% $^\rmi{a}$%

{\em
Faculty of Physics, University of Bielefeld, 
D-33501 Bielefeld, Germany\\}

\vspace*{0.8cm}

\mbox{\bf Abstract}
 
\end{centering}

\vspace*{0.3cm}
 
\noindent
We consider the next-to-leading order thermal production rate 
of heavy right-handed neutrinos in the non-relativistic regime 
$m_\rmi{top} \,\lsim\, \pi T \ll M$, where $m_\rmi{top}$ refers to 
the electroweak scale. Rephrasing the problem in an OPE language and 
making use of different techniques than a previous analysis by Salvio et al, 
we confirm the general structure of their result and many of 
the coefficients. We also extend the analysis to the next order 
in the non-relativistic expansion, thereby revealing the leading 
non-trivial momentum dependence, as well as to NNLO in couplings, 
revealing the leading sensitivity to thermal resummations.  
Our results are expressed as a sum of simple ``master'' structures,
which renders them a suitable starting point for determining the 
next-to-leading order rate also in the relativistic regime $\pi T \sim M$.

\vfill

%\noindent
%PACS numbers: 
%11.10.Wx, %        Finite temperature field theory
%11.15.Ha, %        Lattice gauge theory 
%12.38.Bx, %        Perturbative calculations in QCD
%12.38.Mh, %        Quark--gluon plasma
%14.40.Nd, %        Bottom mesons
%\\
%Keywords:
 
\vspace*{1cm}
  
\noindent
February 2012

\vfill

\end{titlepage}

%%%%%%%%%%%%%%%%%%%%%%%%%%% SECTION %%%%%%%%%%%%%%%%%%%%%%%%%%%%%%%%%%%%%%
%
\section{Introduction}

Weakly interacting 
particles produced by scatterings taking place in a hot plasma
could conceivably play a role in cosmology. For instance, 
some Dark Matter candidates, such as axions, axinos, gravitinos, 
or sterile neutrinos, could originate this way (for reviews see 
e.g.\ refs.~\cite{rev1,rev2}). Another example is the so-called 
Leptogenesis model for explaining the Baryon
Asymmetry of the Universe~\cite{yanagida}, 
in which thermally produced 
out-of-equilibrium right-handed neutrinos
could act as one of the main building blocks. 

Although the Leptogenesis scenario has been thoroughly studied 
(for reviews, see e.g.\ refs.~\cite{lepto1,lepto2}), there are 
some ingredients in the analysis that 
do not appear to be on a sound theoretical footing. In particular, 
it has recently been pointed out that in the ``ultrarelativistic''
regime, meaning temperatures much higher than the mass of
the right-handed neutrinos, $\pi T \gg M \gsim m_\rmi{top}$, 
the expressions that 
had been used are not correct even at leading order in the coupling 
constants (because of infrared sensitivity infinitely many loop orders 
contribute to the leading-order result in this kinematic regime), and
that they had therefore underestimated the production rate~\cite{anisimov}. 
On the other hand, in another recent contribution, it was shown that 
in the ``non-relativistic regime'', $m_\rmi{top} \ll \pi T \ll M$, 
the significance
of loop effects had probably been overestimated in the literature, 
because they had been only 
partly accounted for, whereby an important 
cancellation was missed~\cite{salvio}. (More complicated issues 
like CP-violation could contain even larger 
uncertainties~\cite{cp2}--\cite{cpn}.)

In view of the mentioned developments, it might
be useful to revisit the analysis of the right-handed
neutrino production rate also  in the ``relativistic'' 
regime, $m_\rmi{top} \lsim \pi T \sim M$, checking the importance of 
loop corrections through an explicit computation. Here, we 
take~a first step in this direction, by formulating the setup of 
a next-to-leading order (NLO) analysis, and by carrying it 
out in the non-relativistic regime previously considered 
in ref.~\cite{salvio}. Given that the analysis of ref.~\cite{salvio} 
is exceedingly complicated, our goal was to  reproduce
its results by simpler methods, and also to extend them 
up to a higher order in an expansion in $(\pi T/M)^2$. 

It is important to understand that 
when discussing a non-relativistic regime, we place ourselves
in the position of the particles produced, with 
an invariant mass $\mathcal{K}^2 = M^2 \gg (\pi T)^2$, a kinetic energy 
$k^2 / (2M) \sim \pi T$, and therefore an average velocity 
$k/M \sim \sqrt{\pi T/M} \ll 1$. From the plasma perspective,
however, the four-momentum $\mathcal{K}$ is  
an external probe, which has no dynamical effect. It simply represents
a ``hard'' external scale, with $k^0, k, |k^0\pm k | \gg \pi T$. 
A proper tool for addressing this 
kinematic situation is the Operator Product Expansion~(OPE)~\cite{OPE}, 
formulated within the context of thermal field theory only rather
recently~\cite{simon}, and subsequently applied e.g.\ to 
correlators of the energy-momentum tensor of a Yang-Mills 
plasma~\cite{Bulk_OPE,Shear_OPE}. Understanding
the computation in this language has significant conceptual 
benefits, for instance by leading to a general framework 
for analyzing the infrared (IR) sensitivity of the 
results~\cite{simon}, a topic that remains otherwise 
to be studied empirically~\cite{salvio}.  

%%%%%%%%%%%%%%%%%%%%%%%%%%%%% SECTION %%%%%%%%%%%%%%%%%%%%%%%%%%%%%%%%%%%%
%
\section{Setup}
\la{se:setup}

We consider a plasma made of Standard Model particles, 
in thermal equilibrium at a temperature $T$, and interacting
with right-handed neutrinos, of mass $M$, through Yukawa 
interactions, parametrized by a coupling $h_\nu$. (It would be 
natural to assume the Yukawa couplings to build a $3\times 3$
matrix, but our results are flavour-blind so we can simplify
the notation without a loss of generality.) 
As long as the density of the right-handed
neutrinos is below the equilibrium value, their production
rate can be computed from first principles using a
linear response or Kubo type analysis. Then, essentially, 
the production rate is determined by 
the cut (or imaginary part) of the self-energy of the right-handed
neutrinos.

An explicit derivation 
of the production rate has been presented 
in ref.~\cite{hadronic}.\footnote{%
 The notation in ref.~\cite{hadronic}
 assumed a broken electroweak symmetry, 
 but the derivation goes through also with a dynamical doublet.
 The derivation was extended to $\mu_\ell\neq 0$ 
 in ref.~\cite{shifuller}.
 } 
Letting $\tilde\phi \equiv i \sigma_2\phi^*$;  
denoting by $\ell$ a lepton doublet; 
and by $\aL, \aR$ the left and right projectors, 
$\aL \equiv (1-\gamma_5)/2$, $\aR \equiv (1+\gamma_5)/2$, 
the differential production rate reads
\be
 \frac{{\rm d}N (\mathcal{K})}{{\rm d}^4\mathcal{X} {\rm d}^3\vec{k}}
 = \frac{|h_{\nu\rmii{B}}|^2}{(2\pi)^3 k^0}
   \,
   \tr\Bigl\{ 
     \bsl{\mathcal{K}} \aL 
   \Bigl[ \nF{}(k^0-\mu_\ell) \rho(\mathcal{K}) + 
   \nF{}(k^0 + \mu_\ell) \rho(-\mathcal{K})\Bigr] \aR
   \Bigr\} + \rmO(|h_{\nu\rmii{B}}|^4)
 \;, \la{master1}
\ee
where 
$h_{\nu\rmii{B}}$ is the bare neutrino Yukawa coupling;
$\nF{}(k^0) \equiv 1/[\exp(k^0/T)+1]$ is the Fermi distribution; 
$\mu_\ell$ denotes a leptonic chemical potential;    
$\rho$ is the spectral function related 
to the composite operator
${\tilde\phi}^\dagger{\ell}$; 
$\mathcal{X} = (x^0,\vec{x})$; and
$\mathcal{K} = (k^0,\vec{k})$ is an on-shell four-momentum, 
with $k^0 = \sqrt{k^2 + M^2}$. If the plasma is 
charge-symmetric, i.e.\ $\mu_\ell = 0$, then $\rho$
is symmetric in $\mathcal{K}\to -\mathcal{K}$ and the two 
terms can be combined; we make this assumption
in the following.

For a practical computation, 
it is convenient to employ Euclidean conventions.
We define
\be
 \Pi_E(K) \equiv |h_{\nu\rmii{B}}|^2 \, 
 \tr\Bigl\{ 
   i \bsl{K}  
   \Bigl[
     \int_0^\beta \! {\rm d}\tau \int_\vec{x} e^{i K\cdot X}
   \Bigl\langle
     (\tilde{\phi}^\dagger \aL \ell)(X) \; 
     (\bar{\ell}  \aR \tilde\phi)(0)
   \Bigr\rangle^{ }_T 
   \Bigr]
 \Bigr\}
 \;, \la{PiE_def}
\ee
where now $X = (\tau,\vec{x})$;  
the vector $K = (k_n,\vec{k})$ is Euclidean, 
with $k_n = (2n+1)\pi T$, $n\in\mathbbm{Z}$; 
and $\langle ... \rangle^{ }_T$ refers to 
a thermal expectation value.
Defining a differential ``decay rate'', $\Gamma$,  
in accordance with ref.~\cite{salvio} as 
\be
  \frac{{\rm d}N (\mathcal{K})}{{\rm d}^4\mathcal{X} {\rm d}^3\vec{k}}
 \equiv \frac{2\nF{}(k^0)}{(2\pi)^3 } \; \Gamma(\mathcal{K})
 \;, \la{Gamma_def}
\ee
then \eqs\nr{master1}, \nr{PiE_def} together with the standard relation
between the Euclidean correlator and the spectral function
($\rho = 
\im \Pi_R  = 
\im \{\Pi_E \}_{k_n\to -i [k^0 + i 0^+]}$, 
cf.\ e.g.\ ref.~\cite{lebellac}) imply that
\be
 \Gamma(\mathcal{K}) = \frac{1}{k^0}
 \im\bigl\{ \Pi_E(K) \bigr\}_{k_n \to -i [k^0 + i 0^+]}
 \;. \la{master2}
\ee
Here $\im$ refers to a discontinuity, or cut, across the 
real $k^0$-axis, and the relation is valid even on the 
non-perturbative level.  

To compute the correlator of \eq\nr{PiE_def}, let us establish 
conventions for the relevant part of the Standard Model. 
The Higgs field interacts 
according to the Euclidean Lagrangian
\be
 \mathcal{L}_\phi = (D_\mu\phi)^\dagger(D_\mu\phi)
  + m_\rmii{B}^2 \phi^\dagger\phi
 + \lambda_\rmii{B} (\phi^\dagger\phi)^2 + 
 h_{t\rmii{B}}\, \bar{q}\,  \aR \tilde\phi\, t 
 + 
 h_{t\rmii{B}}^*\, \bar{t}\, \tilde\phi^\dagger \aL q
 \;.  \la{L_phi}
\ee  
The mass parameter, $m_\rmii{B}^2$, will mostly be omitted because we assume 
that $\pi T \gsim m_\rmi{top}$ and, as already alluded to above, 
the NLO results turn out to be IR safe in this 
regime (cf.\ \se\ref{se:OPE}). Technically
the calculation is performed as if we were in the symmetry restored
phase although, as will become clear later on, this assumption can be 
modestly relaxed. When acting on the Higgs the covariant
derivative takes the form 
\be
 [ D_\mu\,\phi ]^{ }_{m} = 
 \Bigl(
   \delta_{mn} \partial_\mu - i g_{2\rmii{B}}  T^a_{mn} A^a_\mu 
   + i g_{1\rmii{B}} T^0_{mn} B^{ }_\mu
 \Bigr)\phi^{ }_n
 \;, 
\ee
where $m,n \in \{1,2\}$; 
$A^a_\mu,B^{ }_\mu$ are 
the SU$_\rmi{L}$(2) and U$_\rmi{Y}$(1) gauge fields, 
respectively; $T^a$ are Hermitean generators of SU$_\rmi{L}$(2), 
normalized as $\tr [T^a T^b] = \fr12 \delta^{ab}$; 
$T^0_{mn} \equiv \fr12 \delta_{mn}$; 
and the couplings appearing are bare ones. 
The quark Yukawa interaction in \eq\nr{L_phi} 
contains the doublet $q = (t \; b)^T$.
The leptons interact according to the Euclidean Lagrangian 
\be
 \mathcal{L}_\ell = 
 \bar{\ell} \gamma_\mu D_\mu \ell 
 \;,  \la{L_ell}
\ee
where $\ell = (\nu \; e)^T$. When acting on the lepton doublet, 
the covariant derivative reads
\be
 [ D_\mu\,\ell\, ]^{ }_m  = 
 \Bigl(
   \delta_{mn} \partial_\mu  
  - i \Bigl[ 
    g_{2\rmii{B}}  T^a_{mn} A^a_\mu +  g_{1\rmii{B}} T^0_{mn} B^{ }_\mu
  \Bigr] \aL - i \Bigl[ g_{1\rmii{B}} 
 \delta_{m2}\delta_{n2} B^{ }_\mu \Bigr] \aR
 \Bigr) \ell^{ }_n
 \;. 
\ee

%%%%%%%%%%%%%%%%%%%%%%%%%%%%% SECTION %%%%%%%%%%%%%%%%%%%%%%%%%%%%%%%%%%%%
%
\section{Issues with regularization}
\la{se:gamma5}

A loop computation in quantum field theory necessitates 
regularization, and by far the most convenient choice for this
is the dimensional one, which we also adopt here. Unfortunately, with
chiral gauge theories dimensional regularization leads to 
inevitable problems. Although well-known, we briefly remark on
some of the issues in this section. 

In $D$ dimensions, the normal (Euclidean) 
Dirac matrices can be taken to satisfy
\be
 \{ \gamma_\mu , \gamma_\nu \} = 2 g_{\mu\nu } \equiv 2\delta_{\mu\nu}
 \;, \quad
 \gamma_\mu^\dagger = \gamma_\mu 
 \;, \quad
 \tr (1) = 4 
 \;. 
\ee
According to the 't Hooft - Veltman~\cite{tHV,bm} convention a Hermitean 
$\gamma_5$ can be defined as 
\be
 \gamma_5 \equiv \gamma_0 \gamma_1 \gamma_2 \gamma_3
 \;, \quad
 \gamma_5^2 = {1}
 \;. \la{tHV_g5}
\ee
An important implication from here is that~\cite{JK,BW} 
\be
 \aR \gamma_\mu \aL = \tilde \gamma_\mu \aL
 \;, \la{problem}
\ee
where 
\be
 \tilde \gamma_\mu \equiv
 \left\{ 
 \begin{array}{rl}
   \gamma_\mu\;, & \mu \le 3 \\
   0\;, & \mu > 3
 \end{array} 
 \right.
 \;.
\ee
In the (free) kinetic part of \eq\nr{L_ell} we might then imagine defining 
a Lagrangian as 
$
 \bar{\ell} \gamma_\mu \partial_\mu \ell \stackrel{?}{\to}
  \bar{\ell} \tilde{\gamma}_\mu \partial_\mu \ell  =
  \bar\ell \aR \tilde{\gamma}_\mu \partial_\mu \aL \ell + 
  \bar\ell \aL \tilde{\gamma}_\mu \partial_\mu \aR \ell
$,
respecting gauge symmetry; but then the propagator is four-dimensional 
and not properly regularized. Or, we could keep the regularized form, 
\ba
 \bar{\ell} \gamma_\mu \partial_\mu \ell  & = & 
  \bar{\ell} (\aR+\aL) {\gamma}_\mu \partial_\mu (\aR+\aL) \ell \nn 
 & = & 
  \bar\ell \aR {\gamma}_\mu \partial_\mu \aL \ell + 
  \bar\ell \aR {\gamma}_\mu \partial_\mu \aR \ell + 
  \bar\ell \aL {\gamma}_\mu \partial_\mu \aL \ell + 
  \bar\ell \aL {\gamma}_\mu \partial_\mu \aR \ell
 \;, 
\ea 
leading to the usual $D$-dimensional propagator; but 
then gauge symmetry is (slightly) broken by the coupling of the 
different chiralities. Though extremely naive, these remarks
already illustrate some of the complications encountered. Still, 
consistent computations are possible if 
counterterms and operator mixing are properly accounted for~\cite{JK,BW}, 
and a closely related practical recipe has also been 
put forward~\cite{SL}.

In the prescription of ref.~\cite{SL}, 
one can conveniently work with $D$\/-dimensional
Dirac matrices after defining axial-vector currents with the structure 
$\gamma_\mu\gamma_5\rightarrow \fr12 [\gamma_\mu,\gamma_5] = 
\frac1{3!}\varepsilon_{\mu\nu\rho\sigma}\,
\gamma_\nu\gamma_\rho\gamma_\sigma$, pulling 
the antisymmetric tensors $\varepsilon$ outside the actual integrals 
and utilizing only  their antisymmetry inside the traces.
The latter is made explicit by writing 
\ba\la{g5Larin}
 \fr12[ \gamma_\mu, \gamma_5] & = &
 \frac1{12}\,\varepsilon_{\mu\nu\rho\sigma}
 \left(\gamma_\nu\gamma_\rho\gamma_\sigma
 -\gamma_\sigma\gamma_\rho\gamma_\nu\right)\;.
\ea
As a further simplification, it has been argued in ref.~\cite{SL}
that a naively anticommuting $\gamma_5$ in combination with 
$\gamma_5^2={1}$ can be used in fermionic traces
containing more than one $\gamma_5$, with the exception of 
closed fermion loops. The consistency of this 
prescription has been verified up to 3-loop order in connection with 
singlet as well as non-singlet axial-vector operators 
in QCD~\cite{SL,SLns}.
We will employ this recipe as a ``minimal crosscheck'', commenting
also on how it differs from the 't Hooft - Veltman scheme 
when closed fermion loops are present. 

It is important to realize, however, that in the computation 
of the spectral function according to \eq\nr{master2} some of 
the ambiguities affecting the Euclidean correlator drop out. 
Indeed, a non-zero cut arises from logarithms or, in the context 
of dimensional regularization, from terms containing $1/\epsilon$-poles
(cf.\ \eq\nr{mom_cut}). So, in terms of the Euclidean correlator, it 
is only necessary to get the $1/\epsilon$-poles correct. Of course, 
the bare parameters also bring along $1/\epsilon$-poles, 
so lower-order graphs need to be worked out 
to a higher depth in the $\epsilon$-expansion.  

%%%%%%%%%%%%%%%%%%%%%%%%%%%%% SECTION %%%%%%%%%%%%%%%%%%%%%%%%%%%%%%%%%%%%
%
\section{Leading order at zero temperature}
\la{se:lo}

In order to get going, we start
by an almost trivial step, performing a leading order analysis
at zero temperature. This has the lucky feature of showing that  
at leading order, we are free from the subtleties  
of \se\ref{se:gamma5} to all orders in $\epsilon$.
 
The starting point is to carry out Wick contractions  in \eq\nr{PiE_def}.
Then, although \eq\nr{problem} implies 
that the external momentum appears as 
\be
 \aR \bsl{K} \aL = \bsl{\tilde{K}} \aL 
 \;, 
\ee
rotational symmetry guarantees that we are free to choose $K$ 
to have {\em at most} $\min(D-1,3)$ non-zero spatial components. 
With this choice, 
$
 \bsl{\tilde{K}} \aL = \bsl{K} \aL
$, 
and 
\be
 \tr[i \bsl{K} \aL (i \bsl{K} - i \bsl{P}) \aR ]
 = 2 \tilde{K} \cdot (\tilde{P} - \tilde{K}) 
 = 2 K \cdot (P - K)
 \;. \la{simpl}
\ee
So we have the same expression as in Naive Dimensional
Regularization (NDR), in which it is assumed, despite 
algebraic inconsistencies, that $\gamma_5$ anticommutes
with all $\gamma_\mu$. 

\vspace{2mm}

With \eq\nr{simpl} at hand, a few steps lead to 
\be
 \Pi_E(K) 
 = 4 |h_{\nu\rmii{B}}|^2 \Tint{P} 
 \frac{K\cdot(P- K)}{P^2(P-K)^2}
 = 2 |h_{\nu\rmii{B}}|^2 \Tint{P} 
 \biggl[ 
   \frac{1}{(P-K)^2} - \frac{1}{P^2} -\frac{K^2}{P^2(P-K)^2}  
 \biggr]\;, \la{PiE_0}  
\ee
where the additional factor 2 comes from the isospin trace, 
and we completed squares in the numerator. 
At zero temperature, with 
$
 \Tinti{P}\to \int_P
$, 
only the last term contributes, and we obtain\footnote{% 
 The $\msbar$ scale parameter is introduced in a usual way,
% $ 
%  \mu^{2} =  \bmu^{2}
%  {e^{\gammaE}}/{4\pi}
% $, 
 inserting 
 $ 
  1 = \mu^{-2\epsilon} \bmu^{2\epsilon}
  \frac{\exp({\gamma_\rmiii{E} \epsilon})}
  {(4\pi)^\epsilon}
 $.
 }  
\be
 \Pi^{(0)}_{E}(K) = -2 |h_{\nu\rmii{B}}|^2 K^2 
 \frac{\mu^{-2\epsilon}}{(4\pi)^2}
 \biggl( \frac{1}{\epsilon} + \ln\frac{\bmu^2}{K^2} + 2 + 
 \rmO(\epsilon) \biggr) 
 \;. \la{vac}
\ee
Subsequently, \eq\nr{master2} yields
\be
 \Gamma^{(0)}_{ }(\mathcal{K}) 
 = \frac{|h_{\nu\rmii{B}}|^2 \mathcal{K}^2}{8\pi k^0} + \rmO(\epsilon)
 \;, \la{Gamma_0_tree}
\ee
where we have denoted 
$\mathcal{K}^2 \equiv (k^0)^2 - k^2$; and  
the Euclidean four-momentum transforms as 
$K^2 = k_n^2 + k^2 \to -\mathcal{K}^2
  - i \mathop{\mbox{sign}}(k^0)\, 0^+$.
The result in \eq\nr{Gamma_0_tree} agrees with 
\eq(3) of ref.~\cite{salvio}.

\vspace{2mm}

% As a consistency check, 
For later reference, let us verify \eq\nr{PiE_0}
using the prescription introduced below \eq\nr{g5Larin}.
Performing the isospin trace, but keeping the Dirac trace 
for the moment, we get
\ba\la{g5PiE_0}
 \Pi_E(K) &=& -2 |h_{\nu\rmii{B}}|^2\, 
 \tr[\gamma_\mu \aL\! \gamma_\nu\, \aR]\, 
 \Tint{P} \frac{K_\mu\,(K-P)_\nu}{P^2(P-K)^2} \,.
\ea
Now, noting that
$
 \tr (\gamma_\mu \aL \gamma_\nu \aR)
 = \fr14 \tr \{\gamma_\mu (\gamma_\nu - \gamma_5 \gamma_\nu \gamma_5)
 + \gamma_\mu [\gamma_\nu,\gamma_5] \} 
$, the commutator part leads to 
$\frac1{24}\,
\varepsilon_{\nu\rho\sigma\kappa}\,
\tr[\gamma_\mu\gamma_\rho\gamma_\sigma\gamma_\kappa
-\gamma_\mu\gamma_\kappa\gamma_\sigma\gamma_\rho]$ and vanishes
once the trace is 
expressed as a product of metric tensors. In the 
first part, the rule of ref.~\cite{SL} with two $\gamma_5$'s 
leads immediately to 
$\tr[\gamma_\mu \aL \gamma_\nu \aR]\to 2g_{\mu\nu}$, 
agreeing with \eq\nr{PiE_0}. In contrast, a strict use of 
\eq\nr{tHV_g5} with two $\gamma_5$'s
leads to the middle equation in \eq\nr{simpl} and 
necessitates a further argument.

%%%%%%%%%%%%%%%%%%%%%%%%%%%%% SECTION %%%%%%%%%%%%%%%%%%%%%%%%%%%%%%%%%%%%
%
\section{General method at non-zero temperature}
\la{se:method}

Proceeding to NLO, we illustrate some details 
by computing explicitly the ``Higgs correction'', i.e.\ terms proportional 
to $\lambda$.

%%%%%%%%%%%%%%%%%%%%%%%%%% SUBSECTION %%%%%%%%%%%%%%%%%%%%%%%%%%%%%%%%%%%%
%
\subsection{Wick contractions}

Carrying out 
the Wick contractions and making use of \eq\nr{simpl} we obtain, 
in analogy with \eq\nr{PiE_0},  
\ba
  \EleB \quad
  \Pi_E^{(\lambda)} & = & 24 |h_{\nu\rmii{B}}|^2 \lambda_\rmii{B} \,
  \Tint{P,Q} \frac{K\cdot(K-P)}{Q^2P^4(K-P)^2} 
   \nn & = & 
  12 |h_{\nu\rmii{B}}|^2 \lambda_\rmii{B} \,
  \Tint{P,Q} \frac{K^2 + (K-P)^2 - P^2}{Q^2P^4(K-P)^2} 
  \nn & = & 
  12 |h_{\nu\rmii{B}}|^2 \lambda_\rmii{B} \,
  \Tint{P,Q} \frac{1}{Q^2}
  \biggl[ 
   \frac{1}{P^4} - \frac{1}{P^2(K-P)^2} + \frac{K^2}{P^4(K-P)^2} 
  \biggr] 
  \;. \hspace*{6mm} \la{lam_1}
\ea
The factorized structures have well-known expressions
in dimensional regularization~\cite{ae}: 
\ba
 \Tint{Q} \frac{1}{Q^2}
 & = &  \int_\vec{q} \frac{\nB{}(q)}{q}
 =
  \frac{T^2 \mu^{-2\epsilon}}{12} 
 \biggl\{ 1 + 2\epsilon 
 \biggl[ 
   \ln\biggl(\frac{\bmu}{4\pi T}\biggr) + 1 + \frac{\zeta'(-1)}{\zeta(-1)}
 \biggr] 
 + \rmO(\epsilon^2) \biggr\}
 % \frac{T^2}{12}  + \rmO(\epsilon)
 \;, \la{Ib1} \\ 
 \Tint{P} \frac{1}{P^4}
 & = &  \frac{1-2\epsilon}{2}\int_\vec{p} \frac{\nB{}(p)}{p^3}
 = \frac{\mu^{-2\epsilon}}{(4\pi)^2} 
   \biggl[\frac{1}{\epsilon}
   + 2\ln\biggl(\frac{\bmu e^{\gammaE}}{4\pi T} \biggr)
   + \rmO(\epsilon)
   \biggr]
  \;, \la{Ib2}
\ea
where $\nB{}(p) \equiv 1/[\exp(p/T)-1]$ denotes the Bose distribution, 
and a partial integration was carried out for obtaining the first 
representation of the latter term.  (It is worth noting that this 
term has a power-like IR divergence which is not visible in 
dimensional regularization; the $1/\epsilon$
has an ultraviolet (UV) origin. Nevertheless, as long as dimensional
regularization is applied consistently, there is no reason to 
worry; we return to this in the paragraph 
following \eq\nr{lam_final}, and a comprehensive analysis is 
presented in \se\ref{se:OPE}.)
However, the product of \eqs\nr{Ib1}, \nr{Ib2} is $K$-independent, 
so there is no cut, and no contribution to $\Gamma(\mathcal{K})$.

%%%%%%%%%%%%%%%%%%%%%%%%%% SUBSECTION %%%%%%%%%%%%%%%%%%%%%%%%%%%%%%%%%%%%
%
\subsection{Single-pole cut}
\la{ss:single}

Non-trivial cuts arise from the 2nd and 3rd terms of \eq\nr{lam_1}. 
Starting with the second term,
the Matsubara sum can be carried out exactly, yielding 
\be
 \Tint{P} \frac{1}{P^2(K-P)^2}
 = 
 \int_P \frac{1}{P^2(K-P)^2} + 
 \int_\vec{p} \frac{\nB{}(p) - \nF{}(p)}{p}
 \biggl[ \frac{1}{(K-P)^2} \biggr]^{ }_{P}
 \;, \la{pre_term1}
\ee
where the notation $[...]^{ }_{P} \equiv \fr12 \sum_{p_n = \pm ip}(...)$
corresponds to setting the thermal line on-shell;  
and the integration variable
was renamed in the fermionic cut. 
The vacuum term has the familiar structure of \eq\nr{vac}, 
with a cut $\frac{1}{16\pi}$, whereas the thermal part can 
formally be expanded in $p$, given that the $p$-integration
is exponentially convergent due to the thermal distributions: 
\be
 \biggl[ \frac{1}{(K-P)^2} \biggr]_{P} = 
 \biggl[ \frac{1}{K^2 - 2 K\cdot P} \biggr]_{P} = 
 \biggl[ \frac{1}{K^2}  + \frac{ 2 K\cdot P}{K^4} 
 + \frac{4 (K\cdot P)^2}{K^6} + \ldots \biggr]_{P}
 \;. \la{ope}
\ee
This yields an OPE-type expansion. However, as long as 
we stay away from the light-cone 
($\re[K^2]_{k_n\to -i[k^0+i0^+]} \neq 0$), this 
expansion has {\em no cut}. Therefore, a contribution
only arises from the vacuum part of \eq\nr{pre_term1},
multiplying \eq\nr{Ib1}:
\be
 \im\biggl\{
   \Tint{P,Q} \frac{1}{Q^2 P^2(K-P)^2}
 \biggr\}_{k_n\to -i[k^0+i0^+]}
 = \frac{1}{16\pi} \frac{T^2}{12} + \rmO(\epsilon)
 \;. \la{term1}
\ee

It is useful for later reference to repeat the analysis 
of the thermal part in a more detailed way. 
For this purpose we set $\vec{k} = \vec{0}$ and 
work out the cut exactly. Given that for $k=0$
\be
 \biggl[ \frac{1}{(K-P)^2} \biggr]_{P;\,k=0} = 
% \biggl[ \frac{1}{ k_n^2 - 2 k_n p_n } \biggr]_{P;\,k=0}  
 \frac{1}{2k_n} \biggl( \frac{1}{k_n - 2 i p} + 
 \frac{1}{k_n + 2 i p} \biggr) = 
 \frac{1}{4ip} \biggl( \frac{1}{k_n - 2 i p} - 
 \frac{1}{k_n + 2 i p} \biggr)
 \;, 
\ee
we get 
\be
 \im\biggl\{ \biggl[ \frac{1}{(K-P)^2}
     \biggr]_{P;\,k=0} \biggr\}_{k_n \to -i [k^0 + i 0_{ }^+] }
  = \frac{\pi}{4p} 
 \Bigl[\delta(k^0 - 2 p) - \delta(k^0 + 2 p) \Bigr]
 \;. \la{delta}
\ee
So, for $k^0 = \sqrt{k^2 + M^2} > 0$,  
\ba
 \im\biggl\{ 
  \int_\vec{p} \frac{\nB{}(p) - \nF{}(p)}{p}
  \biggl[ \frac{1}{(K-P)^2} \biggr]_{P;\, k=0}  
 \biggr\} 
 & = &  \frac{1}{8\pi} \int_0^\infty \! {\rm d}p \, 
   \bigl[\nB{}(p) - \nF{}(p)\bigr]
   \delta(k^0 - 2 p) 
 \nn
 & = & 
  \frac{1}{16\pi} 
   \Bigl[\nB{}\Bigl( \frac{k^0}{2} \Bigr) - 
    \nF{}\Bigl(\frac{k^0}{2}\Bigr)\Bigr]
 \;. \la{single_ex}
\ea
In other words we do find a correction to the 
$\frac{1}{16\pi}$ appearing in \eq\nr{term1} but it is  
exponentially small, suppressed by $e^{-k^0/T}$ 
in the regime $k^0 \gg T$ that we are interested in. 

%%%%%%%%%%%%%%%%%%%%%%%%%% SUBSECTION %%%%%%%%%%%%%%%%%%%%%%%%%%%%%%%%%%%%
%
\subsection{Double-pole cut}
\la{ss:double}

It remains to work out the 3rd term of \eq\nr{lam_1}.
This is slightly less trivial than the 2nd term 
because of the appearance of the double pole. Nevertheless the
basic point is the same: denoting $E_p \equiv p$, 
$E_{kp} \equiv |\vec{k-p}|$, we get
\ba
 & & \hspace*{-1cm} \Tint{P} \frac{1}{P^4(K-P)^2} =  
 \int_\vec{p} \biggl(- \frac{\partial}{\partial E_p^2} \biggr)
  T\sum_{p_n} \frac{1}{(p_n^2 + E_p^2)[(k_n-p_n)^2 + E_{kp}^2]}
 \nn & = & 
 \int_\vec{p} \biggl(- \frac{\partial}{\partial E_p^2} \biggr)
 \biggl\{
   \int_{-\infty}^{\infty} \! \frac{{\rm d}p_0}{2\pi}
   \frac{1}{(p_0^2 + E_p^2)[(k_n-p_0)^2 + E_{kp}^2]} 
  \nn & & \hspace*{1cm} \;
   + \, \frac{\nB{}(E_p)}{E_p} 
     \biggl[ \frac{1}{(k_n - p_n)^2 + E_{kp}^2} \biggr]_{P}
   - \frac{\nF{}(E_{kp})}{E_{kp}} 
     \biggl[ \frac{1}{p_n^2 + E_p^2} \biggr]_{K-P}
 \biggr\} 
 \nn & = & 
 \int_P \frac{1}{P^4 (K-P)^2} + \int_\vec{p} 
 \biggl\{
   - \frac{\partial}{\partial E_p^2} \frac{\nB{}(E_p)}{E_p} 
   + \frac{\nF{}(E_p)}{E_p} \frac{\partial}{\partial E_{kp}^2}  
 \biggr\} 
 \biggl[ \frac{1}{(k_n - p_n)^2 + E_{kp}^2} \biggr]_{P}
 \;, \nn \la{term2_0}
\ea
where we substituted $\vec{p}\to \vec{k-p}$ in the last term. 
The thermal part is more complicated than before, 
%(it can be simplified by writing 
%$\partial^{ }_{E_p} = \frac{\rm d}{{\rm d}p} -
%  \frac{{\rm d}E_{kp}}{{\rm d}p} \partial^{ }_{E_{kp}}$, and carrying 
%out a partial integration with respect to $p$), 
but it is clear that it can again 
be expanded in inverse powers of $1/K^2$, and all resulting
integrals are exponentially convergent in the UV 
(in the IR there are divergences like in \eq\nr{Ib2} but this
only leads to logarithms of $T$, not of $K$). 
Like in \eq\nr{ope}, 
these terms give no cuts. The zero-temperature term, on the other
hand, evaluates in dimensional regularization to 
\be
 \int_P \frac{1}{P^4 (K-P)^2} 
 = 
 -\frac{K^{-2\epsilon}}{\epsilon K^2}
 \frac{\Gamma(1+\epsilon)\Gamma^2(1-\epsilon)}
 {(4\pi)^{2-\epsilon} \Gamma(1-2\epsilon)}
 = -\frac{\mu^{-2\epsilon}}{(4\pi)^2 K^2}
    \biggl(\frac{1}{\epsilon} + \ln\frac{\bmu^2}{K^2}  \biggr)
   + \rmO(\epsilon) 
 \;. 
\ee
Here $1/\epsilon$ is an IR divergence, but the cut 
is IR finite, and can be extracted 
like in \eq\nr{vac}. In total, then, 
\be
 \im\biggl\{ 
   \Tint{P,Q} \frac{K^2}{Q^2 P^4(K-P)^2}
 \biggr\}_{k_n\to -i[k^0+i0^+]}
 = - \frac{1}{16\pi} \frac{T^2}{12} + \rmO(\epsilon)
 \;. \la{term2}
\ee 

For the benefit of a reader nerved by the free use of 
dimensional regularization for handling 
IR divergences, let us repeat the derivation of
\eq\nr{term2} in a more pedantic way, directly in $D=4$,  
avoiding IR divergences. Setting again $\vec{k} = \vec{0}$
for simplicity, and denoting $E_0^2 \equiv p^2 + m_0^2$,
where $m_0^2 > 0$ is an IR regulator of the scalar line,  
we can rewrite \eq\nr{term2_0} as
\ba
 & & \hspace*{-1cm} \Tint{P} \frac{1}{P^4(K-P)^2} =  
  - \lim_{m_0 \to 0} \frac{{\rm d}}{{\rm d} m_0^2} 
 \int_\vec{p} 
  T\sum_{p_n} \frac{1}{(p_n^2 + E_0^2)[(k_n-p_n)^2 + p^2]}
 \nn & = & 
  - \lim_{m_0\to 0} \frac{{\rm d}}{{\rm d} m_0^2} 
 \int_\vec{p} \frac{1}{4 p E_0}
 \biggl\{ 
 \biggl[
   \frac{1}{ik_n+p+E_0} + \frac{1}{-ik_n+p+E_0}
 \biggr] \Bigl[ 1 - \nF{}(p) + \nB{}(E_0) \Bigr]
 \nn 
 & & \hspace*{3cm} \; + \,  
  \biggl[
   \frac{1}{ik_n+p-E_0} + \frac{1}{-ik_n+p-E_0}
  \biggr] \Bigl[ \nF{}(p) + \nB{}(E_0) \Bigr]
  \biggr\} 
 \;. \la{check}
\ea
Setting $k_n\to -i[k^0 + i 0^+]$ and 
taking the cut, the denominators turn into $\delta$-constraints
like in \eq\nr{delta}. If we choose $m_0 < M$, 
only one of them can get realized: 
\ba
 & & \hspace*{-2cm} 
 \im\biggl\{ \Tint{P} 
 \frac{k_n^2}{P^4(K-P)^2}\biggr\}_{k_n \to -i [k^0 + i 0^+]}  
  \nn & & \; = \,  
   \lim_{m_0\to 0} \frac{{\rm d}}{{\rm d} m_0^2} 
 \int_\vec{p} \frac{\pi k_0^2}{4 p E_0}
   \delta(-k^0+p+E_0)
 \Bigl[ 1 - \nF{}(p) + \nB{}(E_0) \Bigr]
 \nn & & \; = \,  
   \lim_{m_0\to 0} \frac{{\rm d}}{{\rm d} m_0^2} 
  \biggl\{
    \frac{k_0^2 - m_0^2}{16\pi}
    \biggl[ 1 - \nF{}\Bigl( \frac{k_0^2-m_0^2}{2k_0} \Bigr)
   +  \nB{}\Bigl( \frac{k_0^2+m_0^2}{2k_0} \Bigr) \biggr] 
  \biggr\}
 \;. \la{double_ex}
\ea
Taking the derivative  we do reproduce 
\eq\nr{term2}, up to exponentially small terms.  

%%%%%%%%%%%%%%%%%%%%%%%%%% SUBSECTION %%%%%%%%%%%%%%%%%%%%%%%%%%%%%%%%%%%%
%
\subsection{Putting everything together}

Inserting finally \eqs\nr{term1}, \nr{term2} into \eq\nr{lam_1}, 
we obtain
\be
 \im \Bigl\{ \Pi_E^{(\lambda)} \Bigr\}_{k_n\to -i[k^0 + i 0^+]}
  = 
 |h_{\nu\rmii{B}}|^2 \lambda_\rmii{B} \,
 \biggl( -\frac{T^2}{8\pi}  \biggr)
 \;. \la{lam_final}
\ee
This corresponds to the Higgs contribution 
in \eqs\nr{pre_result}, \nr{result}.\footnote{%
 If the exponentially small terms  are included,
 then \eqs\nr{term1}, \nr{single_ex}, \nr{double_ex} 
 imply (for $\vec{k} = \vec{0}$)
 \be 
   \im \Bigl\{ \Pi_E^{(\lambda)} \Bigr\} 
   = - \frac{|h_{\nu\rmii{B}}|^2 \lambda_\rmii{B} T^2}{8\pi} 
   \Bigl\{ 1 + \nB{}(\frac{k_0}{2})- \nF{}(\frac{k_0}{2})
       - \frac{k_0}{4} \Bigl[ \nB{}'(\frac{k_0}{2})+ \nF{}'(\frac{k_0}{2}) 
   \Bigr]
   \Bigr\} \;. \la{full}
 \ee
 }

We now return to the issue of IR
divergences. As has been stressed in the introduction, from 
the point of view of the thermal medium the right-handed neutrino
acts as a ``hard probe''; any thermal effects on its
production rate can be understood in the OPE language.
But in the OPE language thermal effects amount to gauge-invariant
condensates developing expectation values~\cite{simon}. 
These, in turn, obtain IR divergences only at higher loop
orders than considered here (this is checked explicitly 
in \se\ref{se:OPE}). 
Therefore it is guaranteed, 
from general principles, that all IR divergences must cancel
in the spectral function, and we are allowed to handle 
them through dimensional regularization, as long as it 
is applied consistently. Of course, the cancellation of 
IR divergences can also be 
checked directly~\cite{salvio}. At the current order, 
the cancellation in fact probably takes place in every ``master''  
spectral function separately, and for an arbitrary
value of $k^0/T$, as has been observed 
in the fully bosonic case~\cite{Bulk_wdep}. 

Having thereby concluded the discussion of Higgs corrections, 
we note that similar methods work in all other cases as well. 
The full list of structures appearing can be found in appendix~A, 
and the corresponding spectral functions are given in appendix~B.
Whereas for the current paper the 
cuts of appendix~B are sufficient, in some  
contexts  it may 
be desirable to also know the ultraviolet expansions in 
the full Euclidean domain (cf.\ e.g.\ \se\ref{se:OPE}); 
these are listed in appendix~C, 
and have been derived with methods explained 
in refs.~\cite{Bulk_OPE,Shear_OPE}.

%%%%%%%%%%%%%%%%%%%%%%%%%%%%% SECTION %%%%%%%%%%%%%%%%%%%%%%%%%%%%%%%%%%%%
%
\section{Next-to-leading order analysis}
\la{se:nlo}

Including now also contributions from the top quark and from
gauge bosons, we turn to our full NLO expressions. In order to 
obtain a ``universal'' representation, we make use of 
completions of squares and substitutions of integration variables
in order to express the results in terms of a minimal number of
independent ``master'' sum-integrals, listed in appendix~A. 
We specify graph-by-graph results in NDR in terms of 
these masters.\footnote{%
 We have checked that the same results are obtained, 
 for every diagram, with the recipe described below \eq\nr{g5Larin}, 
 whereas in the strict 't Hooft - Veltman scheme there are
 additional terms; 
 cf.\ appendix~D.
 } 
Gauge parameter independence (with respect to both gauge
groups) has been checked separately, so here only the 
Feynman gauge results are shown. They read
\ba
%  \EleA & = & 2 |h_{\nu\rmii{B}}|^2 
%  \Bigl(
%   -\mathcal{J}^{ }_\rmi{a} 
%   +\widetilde{\mathcal{J}}_\rmi{a} 
%   -\mathcal{J}^{ }_\rmi{b}
%  \Bigr)
%  \;, \la{EleA} \\[4mm] 
  \EleB & = & 12 |h_{\nu\rmii{B}}|^2 \lambda_\rmii{B}
  \Bigl(
   -\mathcal{I}^{ }_\rmi{b}
   +\mathcal{I}^{ }_\rmi{c}
   +\mathcal{I}^{ }_\rmi{d}
  \Bigr) 
  \;, \la{EleB} \\[4mm] 
  \EleC & = & 2 |h_{\nu\rmii{B}}|^2 |h_{t\rmii{B}}|^2 \Nc 
  \Bigl(
   2\, \widetilde{\mathcal{I}}_\rmi{b}  
  - 2\, \widetilde{\mathcal{I}}_\rmi{c}  
  - 2\, \widetilde{\mathcal{I}}_\rmi{d}
  +  \widetilde{\mathcal{I}}_\rmi{e}  
  -  \widetilde{\mathcal{I}}_\rmi{f}  
  +  \widetilde{\mathcal{I}}_\rmi{h}    
  \Bigr) 
  \;, \\[4mm] 
  \EleD & = & |h_{\nu\rmii{B}}|^2 (g_{1\rmii{B}}^2 + 3 g_{2\rmii{B}}^2)
  \Bigl[  
    \frac{D}{2} \Bigl(
      -\mathcal{I}^{ }_\rmi{b}
      +\mathcal{I}^{ }_\rmi{c}
      +\mathcal{I}^{ }_\rmi{d}     
    \Bigr)
  \Bigr]
  \;, \\[4mm] 
  \EleE & = & |h_{\nu\rmii{B}}|^2 (g_{1\rmii{B}}^2 + 3 g_{2\rmii{B}}^2)
  \Bigl[  
    \frac{1}{2} \Bigl(
      \mathcal{I}^{ }_\rmi{b}
      -\mathcal{I}^{ }_\rmi{c}
      -\mathcal{I}^{ }_\rmi{d}     
    \Bigr) 
      -\mathcal{I}^{ }_\rmi{e}
      +\mathcal{I}^{ }_\rmi{f}
      -\mathcal{I}^{ }_\rmi{h}         
  \Bigr]
  \;, \\[4mm] 
  \EleF & = & |h_{\nu\rmii{B}}|^2 (g_{1\rmii{B}}^2 + 3 g_{2\rmii{B}}^2)
  \Bigl[  
    \frac{D-2}{2} \Bigl(
      \mathcal{I}^{ }_\rmi{b}
      -\widetilde{\mathcal{I}}^{ }_\rmi{b}
      +\overline{\mathcal{I}}^{ }_\rmi{c}
      -\widehat{\mathcal{I}}^{ }_\rmi{c}
      +\widehat{\mathcal{I}}^{ }_\rmi{d}     
      -\overline{\mathcal{I}}^{ }_\rmi{d}     
      +\widehat{\mathcal{I}}^{ }_\rmi{h'}     
    \Bigr)
  \Bigr]
  \;, \hspace*{1cm} \\[4mm] 
  \EleG & = & |h_{\nu\rmii{B}}|^2 (g_{1\rmii{B}}^2 + 3 g_{2\rmii{B}}^2)
  \Bigl(
      \widetilde{\mathcal{I}}^{ }_\rmi{e}
     - \mathcal{I}^{ }_\rmi{f}
     + \mathcal{I}^{ }_\rmi{g}
     - \mathcal{I}^{ }_\rmi{h}
     - 2\,\widehat{\mathcal{I}}^{ }_\rmi{h}     
     + \mathcal{I}^{ }_\rmi{j}
  \Bigr)
  \;, \la{EleG}
\ea
with dashed, solid, doubled, and wiggly lines representing scalars, 
leptons, quarks, and gauge bosons, respectively. 

Inserting the cuts, or spectral functions, 
from appendix~B; setting $D = 4-2\epsilon$; 
and renormalizing according to 
\ba
 |h_{\nu\rmii{B}}|^2 & = &  |h_\nu(\bmu)|^2 \mu^{2\epsilon} 
 \,\mathcal{Z}_\nu \;, \quad {\rm with} \la{hnuB}\\
  \mathcal{Z}_\nu & \equiv & 
% \; \biggl\{
  1 +  \frac{1}{(4\pi)^2\epsilon}
   \Bigl[
     |h_t|^2 \Nc - \fr34 (g_1^2 + 3 g_2^2) 
   \Bigr] + \rmO(g^4)
% \biggr\} 
   \;,  \la{Znu}
% \\ 
%  |h_{t\rmii{B}}|^2 & = & |h_t(\bmu)|^2 \mu^{2\epsilon} \, 
%  \Bigl\{ 1 + \rmO(g^2) \Bigr\} \;, 
\ea
where $g^2$ denotes a generic renormalized coupling, we obtain 
\ba
 \im \Pi_E(\mathcal{K}) & = & 
 \frac{|h_\nu(\bmu)|^2 \mathcal{K}^2}{8\pi}
 \biggl\{
  1 - \frac{12\lambda}{\mathcal{K}^2} \int_\vec{p} \frac{\nB{}}{p}
 \nn & & \; - \, 
 |h_t|^2\Nc 
 \biggl[
   \frac{1}{(4\pi)^2} 
     \biggl( 
      \ln\frac{\bmu^2}{\mathcal{K}^2} + \fr72
     \biggr)
  + \frac{k_0^2 + k^2/3}{\mathcal{K}^6} \int_\vec{p} \frac{4 p\, \nF{}}{3}
 \biggr]
 \nn & & \; + \, 
 (g_1^2 + 3 g_2^2)
 \biggl[
   \frac{3}{4(4\pi)^2} 
     \biggl( 
      \ln\frac{\bmu^2}{\mathcal{K}^2} + \fr{29}6
     \biggr)
  + \frac{k_0^2 + k^2/3}{\mathcal{K}^6} \int_\vec{p} 
  \frac{p\, (17 \nF{} - 16 \nB{}) }{3}
 \biggr]
 \nn & & \; + \, \rmO\Bigl(g^4,
 \frac{g^3 T^2}{\mathcal{K}^2},
 \frac{g^2 T^6}{\mathcal{K}^6} \Bigr)
 \biggr\}  
 \;. \la{pre_result}
\ea
We note in passing that 
the structure $k_0^2 + k^2/3$ originates from terms like
the last one in \eq\nr{ope}, which after averaging over the 
directions of $\vec{p}$ yields 
\be
 \int\! {\rm d}\Omega_\vec{p}\, [(K\cdot P)^2]^{ }_{P} =  
 p^2 \Bigl( \frac{k^2}{3-2\epsilon} - k_n^2 \Bigr) 
 \;.
\ee
A physical interpretation for this structure is given 
after \eq\nr{traceless}.

%%%%%%%%%%%%%%%%%%%%%%%%%%%%% SECTION %%%%%%%%%%%%%%%%%%%%%%%%%%%%%%%%%%%%
%
\section{Next-to-leading order results}

If we choose 
the renormalization scale as $\bmu = M$ in \eq\nr{pre_result}; 
denote the corresponding renormalized coupling by $|h_\nu|^2$; 
set $\Nc = 3$; and insert \eq\nr{Ib1} together with  
\be
% \int_\vec{p} \frac{\nB{}}{p} = \frac{T^2}{12}
% \;, \quad 
 \int_\vec{p} p\, \nB{} = \frac{\pi^2T^4}{30}
 \;, \quad
 \int_\vec{p} p\, \nF{} = \frac{7\pi^2T^4}{240}
 \;, 
\ee
then the ``decay rate'' from \eq\nr{Gamma_def} becomes
\ba
 \Gamma^{ }_{ }(\mathcal{K}) \!\! & = & \!\! 
 \frac{|h_\nu|^2\, M^2}{8\pi \sqrt{k^2 + M^2}}
 \biggl\{
   1 - \frac{\lambda T^2}{M^2} 
  % \nn & &
   \, - \, 
   |h_t|^2 
   \biggl[      
     \frac{21}{2(4\pi)^2} + 
     \frac{7\pi^2 }{60}
     \biggl( \frac{T^4}{M^4} + \fr43 \frac{k^2 T^4}{M^6} \biggr)
   \biggr]
   \nn & & \hspace*{0.6cm} \; + \, 
   (g_1^2 + 3 g_2^2)
   \biggl[      
     \frac{29}{8(4\pi)^2} - 
     \frac{\pi^2 }{80}
     \biggl( \frac{T^4}{M^4} + \fr43 \frac{k^2 T^4}{M^6} \biggr)
   \biggr]
   % \nn & & 
    \, + \, \rmO\Bigl(g^4,\frac{g^3T^2}{M^2},\frac{g^2T^6}{M^6} \Bigr)
    \biggr\}
 \;. \hspace*{1cm} \la{result}
\ea
This constitutes our main result.
(The $\rmO(g^3)$ correction is worked out in \se\ref{se:OPE} and amounts
to $\lambda T^2 \to \lambda T (T - 3 m_\rmii{H} / \pi)$, where $m_\rmii{H}$
is the thermal mass parameter given in \eq\nr{mH}.)

Equation~\nr{result} can be compared with ref.~\cite{salvio}.
We completely agree on all $T$-independent terms, 
as well as on the fact that
gauge corrections lead to no term proportional to $T^2$. As far as
the Higgs contribution is concerned, we find a result larger by
a factor~2. As far as the top correction is concerned, we find 
that all terms of $\rmO(T^2)$ cancel like in 
gauge corrections, whereas in ref.~\cite{salvio} only a partial 
cancellation was observed, so that a term of $\rmO(T^2)$ was left over. 
The corrections of $\rmO(T^4)$ in \eq\nr{result} were not
considered in ref.~\cite{salvio}. 

As shown in \se\ref{se:method}, 
we have cross-checked the Higgs contribution by independently 
computing the full rate for a general $\pi T / M$ 
and taking the non-relativistic limit only afterwards, 
cf.\ \eq\nr{full}. 
As far as we can judge, the Higgs 
correction in ref.~\cite{salvio} was inferred from the effect
that a thermal mass would have on a vacuum-like result; 
but since there are no IR issues at NLO, results 
emerge from momenta $p  \gg gT$, and
must be insensitive to thermal masses. 
In fact, taking a zero-temperature decay rate
$\sim M^2 - m_0^2$, with $m_0^2$ denoting a vacuum mass, 
and denoting by $m_\rmii{H}^2 = \frac{\lambda T^2}{2}$ the thermal 
mass, then secs.~\ref{ss:single} and \ref{ss:double} 
show that there are two 
contributions, amounting to 
$M^2 - m_0^2 \to M^2 - m_0^2 - m_\rmii{H}^2 - m_\rmii{H}^2 
\frac{{\rm d}}{{\rm d} m_0^2} m_0^2$, rather than a single mass shift
$\exp(m_\rmii{H}^2 \frac{{\rm d}}{{\rm d} m_0^2} )(M^2 -  m_0^2)$
as was assumed in ref.~\cite{salvio}.  One of the contributions
comes from a ``double-pole'' cut and requires a careful analysis. 

Let us finally consider the total rather than the differential
production rate, obtained according to \eq\nr{Gamma_def} as 
\be
 \gamma(T)
 \equiv  \frac{{\rm d}N}{{\rm d}^4\mathcal{X}}
 = \int \! \frac{ {\rm d}^3\vec{k} }{(2\pi)^3 }
 \; {2\nF{}(k^0)} \; \Gamma(\mathcal{K})
 \;.
\ee
A straightforward integration with $\Gamma$ from \eq\nr{result} yields
\ba
 \gamma(T) & = &    
 \frac{|h_\nu|^2\, M^3 T}{(2\pi)^3}
 \biggl\{
   \bigl( 1 + c_1 \bigr) K_1\Bigl( \frac{M}{T} \Bigr)
   + \frac{3 c_2 T}{M} K_2\Bigl( \frac{M}{T} \Bigr)
%  \nn & & \hspace*{3cm} 
 +  \rmO\Bigl[ ({\textstyle\frac{T}{M}})^{\fr12}
  e^{-\fr{2M}{T}} \Bigr]
 \biggr\} 
 \;, \hspace*{1.0cm} \la{gammaT}
\ea
with the coefficients 
\ba
 c_1  & = & 
  - \frac{\lambda T^2}{M^2} 
   \, - \, 
   |h_t|^2 
   \biggl[      
     \frac{21}{2(4\pi)^2} + 
     \frac{7\pi^2 T^4 }{60 M^4}
   \biggr]
    \, + \, 
   (g_1^2 + 3 g_2^2)
   \biggl[      
     \frac{29}{8(4\pi)^2} - 
     \frac{\pi^2 T^4}{80M^4}
   \biggr]
  \;, \\  
 c_2 & = & 
   \, - \, 
   |h_t|^2 
      \frac{7\pi^2 T^4 }{45 M^4}
     \, - \, 
   (g_1^2 + 3 g_2^2)
     \frac{\pi^2 T^4}{60M^4}
  \;. \hspace*{1cm} \la{total_result}
\ea
Uncertainties are like in \eq\nr{result}, except that $c_2$
has only thermal corrections.  
Embedding \eq\nr{gammaT} in cosmology, the number density, 
$n(T) \equiv \frac{{\rm d}N}{{\rm d}^3\vec{x}}$, conventionally
normalized to the total entropy density, $s(T)$, evolves as\footnote{
 The speed of sound squared, $c_s^2$, is often approximated as $\fr13$, 
 which is true in a conformally symmetric situation, but not
 when mass thresholds are crossed or effects from running couplings
 are taken into account. 
 } 
\be
  T\, \frac{{\rm d}}{{\rm d} T} \biggl(\frac{n(T)}{s(T)} \biggr)
 = - \frac{\gamma(T)}{3 c_s^2(T) s(T) H(T) }
 \;, \la{kin_tot_3}
\ee
where $H(T)$ is the Hubble parameter.  This  
equation is correct as long as the differential abundance 
remains below its equilibrium value at each $\vec{k}$ separately, 
cf.\ e.g.\ ref.~\cite{shifuller}; otherwise a ``back reaction'' from 
Pauli exclusion needs to be included. 

%%%%%%%%%%%%%%%%%%%%%%%%%%%%% SECTION %%%%%%%%%%%%%%%%%%%%%%%%%%%%%%%%%%%%
%
\section{OPE representation}
\la{se:OPE}

In \se\ref{se:method}, it was mentioned that if the Euclidean 
correlator $\Pi_E^{ }$ is considered, then some of the individual 
terms appear to be IR divergent. As pointed out in ref.~\cite{simon}, 
the nature of these divergences can be understood by representing 
the result in an OPE form.

We note first that, carrying out a naive dimensionally regularized NLO
computation, the (bare) condensate $\langle \phi^\dagger \phi \rangle^{ }_T$
reads
\be
 \langle \phi^\dagger \phi \rangle^{\rmi{naive}}_T 
 = 
 2 \, \mathcal{J}^{ }_\rmi{a}
  - 12 \lambda_\rmii{B} \, \mathcal{I}^{ }_\rmi{c}
 + 2 |h_{t\rmii{B}}|^2 \Nc \, (  2\, \widetilde{\mathcal{I}}_\rmi{c}  
  -  \widetilde{\mathcal{I}}_\rmi{e} )
  + (g_{1\rmii{B}}^2 + 3 g_{2\rmii{B}}^2) \Bigl(     \frac{1- D}{2}\,
      \mathcal{I}^{ }_\rmi{c}
      +\mathcal{I}^{ }_\rmi{e}     
  \Bigr)  
%  2 \int_\vec{p} \frac{\nB{}(p)}{ p}
% - 2 |h_t|^2\Nc  
% \, (1-2\epsilon)\! \int_\vec{p,q} \frac{\nF{}(p)\nB{}(q)}{pq^3}
% \nn & & \; 
% - \, \frac{24\lambda + (g_1^2 + 3 g_2^2)(3-2\epsilon)}{4}
% \, (1-2\epsilon)\! \int_\vec{p,q} \frac{\nB{}(p)\nB{}(q)}{pq^3}
 + \rmO(g^4) 
 \;. \la{pdp_1}
\ee
Here appear the same IR divergent structures 
as were encountered in \se\ref{se:method}. 
If we now write $\Pi_E^{ } = \Pi_E^{(0)} + \Pi_E^{(T)}$, where $\Pi_E^{(0)}$
denotes the vacuum part, then by inserting the expansions from appendix~C
into \eqs\nr{EleB}--\nr{EleG} and renormalizing according to \eq\nr{hnuB},
the thermal part can be expressed as 
\be
 \Pi_E^{(T)}(K) = -2 |h_\nu|^2 
 \biggl[
   1 + \frac{3\lambda}{8\pi^2}
   \biggl( \ln\frac{\bmu^2}{K^2} + 1 + \rmO(\epsilon) \biggr) + \rmO(g^4)
 \biggr] \mathcal{Z}^{ }_{m}\, \mu^{2\epsilon}
 \langle \phi^\dagger \phi \rangle_T^{\rmi{naive}}
 + \rmO\Bigl(\frac{T^4}{K^2} \Bigr)  
 \;, \la{PiET_ope}
\ee
where 
\be 
 \mathcal{Z}_m \equiv 
 \mathcal{Z}_\nu \biggl[1+\frac{3\lambda}{8\pi^2\epsilon}+\rmO(g^4)\biggr]
 = 1 + \frac{1}{(4\pi)^2\epsilon}
 \biggl[
   6 \lambda + |h_t|^2\Nc - \fr34 (g_1^2 + 3 g_2^2) 
 \biggr] + \rmO(g^4)
 \la{Zm}
\ee
happens to be the renormalization factor related to the Higgs mass parameter, 
$m_\rmii{B}^2 = m_0^2 \mathcal{Z}_m$. 
The prefactor in \eq\nr{PiET_ope}  is IR-safe 
(temperature independent)
as is typical of a Wilson coefficient; all IR-sensitive terms
of $\rmO(T^2)$ have been eaten up by the condensate. 

In order to compute the condensate {\em correctly} at $\rmO(g^2)$
also in the IR, 
we need to add to \eq\nr{pdp_1}, which in this naive form 
represents an ultraviolet contribution
from ``hard'' momenta $p \gsim \pi T$, the contribution from ``soft'' 
momenta, in particular from Matsubara zero modes. This type of computations
are best formulated within dimensionally reduced
effective field theories. In fact the result can be 
extracted e.g.\  from ref.~\cite{pert}, and reads 
\ba
 \langle \phi^\dagger\phi\rangle^{(n=0)}_T 
 & = &  -\frac{m_\rmii{H} T}{2\pi} +
 \frac{T^2}{(4\pi)^2}
 \biggl[ 6 \lambda +
  \frac{g_1^2 m_\rmii{D1}^{ }+ 3 g_2^2 m^{ }_\rmii{D2}}{4 m_\rmii{H}}
 \nn & & \hspace*{3cm}
 + \, (g_1^2 + 3 g_2^2)
 \biggl( \frac{1}{4\epsilon} + \ln\frac{\bmu}{2 m_\rmii{H}} + \fr14 \biggr)
 \biggr] + \rmO(g^3)
 \;, \la{pdp_2}
\ea
where the Standard Model thermal masses read, for $\Nc = 3$, 
\be
 m_\rmii{H}^2 =  m_0^2 + 
 \biggl(\frac{\lambda}{2} + \frac{|h_t|^2}{4} + 
 \frac{g_1^2 + 3 g_2^2}{16} \biggr) T^2
 \;, \quad
 m_\rmii{D1}^2 = \frac{11}{6} g_1^2 T^2
 \;, \quad
 m_\rmii{D2}^2 = \frac{11}{6} g_2^2 T^2
 \;. \la{mH}
\ee
Summing together \eqs\nr{pdp_1} and \nr{pdp_2}, multiplying
by $\mathcal{Z}_m$ as required by \eq\nr{PiET_ope}, and for 
convenience also resumming the Debye mass 
contributions of $\rmO(g^3)$ into $m_\rmii{H}^2$, we get
\ba
 \mathcal{Z}_m \langle \phi^\dagger \phi \rangle^{ }_T & = & 
  \frac{T^2}{6} - \frac{T^2}{2\pi}
  \sqrt{ \frac{m_\rmii{H}^2}{T^2} - 
 \frac{ g_1^2 m_\rmii{D1}^{} + 3 g_2^2 m_\rmii{D2}^{ }}
  {16\pi T} }
  \nn & & \; + \, 
  \frac{T^2}{48\pi^2}
  \biggl\{
     -6 \lambda \biggl[ \ln\biggl(
   \frac{\bmu e^{\gammaE}}{4\pi T}\biggl)
     - 3 \biggr]
     - |h_t|^2 \Nc \ln\biggl(
      \frac{\bmu e^{\gammaE}}{8\pi T}\biggl)
  \nn & & \; + \, 
  \frac{3(g_1^2 + 3 g_2^2)}{4}
  \biggl[ 
    \ln\biggl(\frac{\bmu e^{\gammaE}}{4\pi T}
    \biggl) - \fr23 - 2 \gammaE - 
    \frac{2\zeta'(-1)}{\zeta(-1)} + 4 \ln 
   \biggl(\frac{2\pi T}{m_\rmii{H}}\biggl)
  \biggr]
  \biggr\} + \rmO(g^3) 
  \;.  \nn
 \la{pdp_3}
\ea 
Apart from \eqs\nr{Ib1}, \nr{Ib2}, the fermionic 
$
 \int_{\vec{p}}  \frac{\nF{}(p)}{p}  =  \frac{T^2 \mu^{-2\epsilon}}{24}
 \bigl\{ 1 + 2\epsilon 
 \bigl[ 
   \ln\frac{\bmu}{8\pi T} + 1 + \frac{\zeta'(-1)}{\zeta(-1)} 
 \bigr] 
% + \rmO(\epsilon^2)
 \bigr\}
$
was needed here. 
All ultraviolet divergences have nicely cancelled out in \eq\nr{pdp_3}. 

The thermal correction to the ``decay rate'' of \eq\nr{Gamma_def} 
now comes from the cut of the Wilson coefficient in 
\eq\nr{PiET_ope}. It is given by 
\be
 \Gamma^{(T)}(\mathcal{K}) 
 = -\frac{3|h_\nu|^2  }{4\pi \sqrt{k^2 + M^2} }
 \, \bigl[ \lambda + \rmO(g^4) \bigr] 
 \; \mathcal{Z}_m \langle \phi^\dagger \phi \rangle^{ }_T 
 + \rmO(|h_\nu|^2 g^2 T^4)
 \;.
 \la{rate_np}
\ee
Looking back at \eq\nr{pdp_3}, we observe that 
there is a subleading $\rmO(g^3)$ contribution to the decay rate 
(from the term $-\frac{m_\rmii{H} T}{2\pi}\sim \rmO(g)$ in 
$\mathcal{Z}_m \langle \phi^\dagger\phi \rangle^{ }_T$) which is 
sensitive to thermal resummations, however in a computable way. 

Let us finally briefly remark on the terms of $\rmO(T^4)$ in 
\eq\nr{rate_np}. As was pointed out 
in ref.~\cite{simon}, these contain expectation 
values of various components of the energy-momentum tensor. At finite
temperatures, the most important contribution turns out to emerge
from its {\em traceless} part: denoting 
$
 \langle \Theta^{\mu\nu} \rangle^{ }_T = \mathop{\mbox{diag}}(e,p,p,p)
$, 
we may define
\ba
 \langle \hat \Theta^{\mu\nu} \rangle^{ }_T & \equiv &
  \langle \Theta^{\mu\nu} 
 - \fr14 \eta^{\mu\nu} {\Theta^\alpha}_{\alpha} \rangle^{ }_T 
  = \mathop{\mbox{diag}}(e,p,p,p) - \fr14
   \mathop{\mbox{diag}}(\mbox{$+$$-$$-$$-$})(e-3p)
 \nn 
 & = &  
 \fr34 (e+p) \, \mathop{\mbox{diag}}(1,\tfr13,\tfr13,\tfr13)
 \;. \la{traceless}
\ea
Since $e+p = T s$, where $s$ is the entropy density, this 
condensate vanishes at zero temperature. The non-trivial momentum 
dependence in \eq\nr{pre_result} comes from terms
of the type 
$
 \sum_i 
 \mathcal{K}_\mu \mathcal{K}_\nu \langle
   \hat\Theta_i^{\mu\nu} \rangle^{ }_T / 
 \mathcal{K}^6
$, 
just like in ref.~\cite{simon}, where $\hat\Theta_i^{\mu\nu}$ denote
various gauge-invariant subparts of the full tensor (we have not 
worked out the decomposition because it is not needed here). 
Thermal resummations affect $e$ and $p$ 
first at $\rmO(g^3T^4)$~\cite{kapusta}, 
and thus $\Gamma^{(T)}(\mathcal{K})$ at $\rmO(g^5T^4)$.

%%%%%%%%%%%%%%%%%%%%%%%%%%%%% SECTION %%%%%%%%%%%%%%%%%%%%%%%%%%%%%%%%%%%%
%
\section{Conclusions}
\la{se:future}

The purpose of this paper has been to compute the thermal 
production rate of right-handed neutrinos in the non-relativistic
regime, meaning at temperatures much below their mass, but still 
higher than the electroweak scale: $m_\rmi{top}\, \lsim\, \pi T \ll M$. 
(Equivalently, one can speak of a decay rate, 
cf.\ \eq\nr{master2}.)
In this regime, the results
can be organized in the form of an Operator Product Expansion, 
with successively higher powers of $\pi T$ representing thermal
expectation values of gauge-invariant condensates~\cite{simon}.
We have worked at NLO in the coupling constants
(even at NNLO in \se\ref{se:OPE}), 
and to the third order in an expansion in $(\pi T/M)^2$; 
the main result is shown in \eq\nr{result}. 
At order $(\pi T/M)^0$ 
we fully confirm previous NLO results
in the literature~\cite{salvio}, if employing naive 
dimensional regularization for handling $\gamma_5$; 
at order $(\pi T/M)^2$
we confirm the absence of corrections from gauge bosons, 
but find different results for corrections from the Higgs scalar
and the top quark. Our results of $(\pi T/M)^4$ are 
new, and display the leading non-trivial
dependence on the spatial momentum (the terms proportional
to $k^2$ in \eq\nr{result}, which contribute 
at $(\pi T/M)^5$ to the total production rate). 

Despite small differences,
the numerical magnitude of our corrections is similar to 
what was found in ref.~\cite{salvio}. Even though 
our NLO thermal Higgs correction is larger by a factor 2, 
we also find an NNLO term, cf.\ \eq\nr{pdp_3}, which 
numerically cancels about half of the Higgs correction. 
The most substantial difference is that we find a complete
cancellation of top corrections at $\rmO(\pi T/M)^2$.
Therefore our thermal corrections are {\em negative}, 
cf.\ \eqs\nr{result} and \nr{rate_np}, not positive
as was found for the top correction in ref.~\cite{salvio}. 
Their precise numerical
influence on leptogenesis computations 
remains to be inspected but probably the effects
are no larger than was found in ref.~\cite{salvio} 
because the largest (top) term is absent at $\rmO(\pi T/M)^2$.

In scenarios of TeV scale leptogenesis, the rate that we have computed 
could be relevant down to temperatures 
around the electroweak scale. Remarkably, 
\eq\nr{rate_np} shows that then the rate could be extracted
non-perturbatively from Euclidean lattice simulations of the
type that were developed for studying the electroweak phase transition
in the 1990s. 
In addition \eq\nr{rate_np} shows that terms of $\rmO(\pi T/M)^2$
are necessarily proportional to $\lambda$; this can be traced
back to a mismatch of the renormalization factors of the neutrino
Yukawa coupling (\eq\nr{Znu}) and the Higgs mass parameter (\eq\nr{Zm})
which ultimately leads to the logarithm in \eq\nr{PiET_ope}.

As an outlook, we envisage that taking the representations in 
\eqs\nr{EleB}--\nr{EleG} as starting points, it is 
a feasible if hard task to extend the NLO analysis to 
the relativistic regime, $\pi T\sim M$. Indeed a similar 
step has previously been taken in QCD, by going from 
OPE-type results in ref.~\cite{Bulk_OPE} to relativistic
results in ref.~\cite{Bulk_wdep}. We believe that in 
the current context it would be important to keep the
spatial momentum $\vec{k}$ different from zero, 
which makes the analysis more demanding; still,  in principle
similar techniques should work.  
 
%%%%%%%%%%%%%%%%%%%%%%%%% SECTION %%%%%%%%%%%%%%%%%%%%%%%%%%%%%%%%%%%%%
%
\section*{Acknowledgements}

 This work was partly supported by the BMBF under project
% {\em Heavy Quarks as a Bridge between
%      Heavy Ion Collisions and QCD}.
 06BI9002.
 M.L.\ thanks \linebreak Dietrich B\"odeker for useful discussions. 

%%%%%%%%%%%%%%%%%%%%%%%%% SECTION %%%%%%%%%%%%%%%%%%%%%%%%%%%%%%%%%%%%%
%
\section*{Note added}

In the revised version (v3) and Erratum of ref.~\cite{salvio},
the results have been corrected and now agree with ours.

%%%%%%%%%%%%%%%%%%%%%%% APPENDIX %%%%%%%%%%%%%%%%%%%%%%%%%%%%%%%%%%%
%
\appendix
\renewcommand{\thesection}{Appendix~\Alph{section}}
\renewcommand{\thesubsection}{\Alph{section}.\arabic{subsection}}
\renewcommand{\theequation}{\Alph{section}.\arabic{equation}}

%%%%%%%%%%%%%%%%%%%%%%%%%%%%%% SECTION %%%%%%%%%%%%%%%%%%%%%%%%%%%%%%%%%
%
\section{Definitions of master sum-integrals}
\la{app:A}

Denoting by $P,Q$ bosonic and by $K,R,S$ fermionic
Matsubara four-momenta, and employing the usual conventions
$\Tinti{P}$ and $\Tinti{\{R\}}$ for the corresponding 
measures, the master sum-integrals entering
the computation are defined as follows (in the graphical notation, 
a dashed line indicates a bosonic propagator, a solid line
a fermionic one, a filled blob a squared propagator, 
and a cross that the momentum 
appears in the numerator as well):
\ba
 \TopoST(\Lqu,\Asc) \quad
 \mathcal{J}^{ }_\rmi{a} & \!\!\equiv\!\! & 
 \Tint{P} \frac{1}{P^2}
 \;,
 \\
 \TopoST(\Lqu,\Aqu) \quad
 \widetilde{\mathcal{J}}^{ }_\rmi{a} & \!\!\equiv\!\! & 
 \Tint{\{R\}} \frac{1}{R^2}
 \;,
 \\ 
 \TopoSB(\Lqu,\Asc,\Aqu) \quad
 \mathcal{J}^{ }_\rmi{b} & \!\!\equiv\!\! & 
 \Tint{P} \frac{K^2}{P^2(P-K)^2}
 \;,
 \\
 \TopoST(\Lqu,\Asc) \times
 \TopoSB(\Lqu,\Asc,\Aqu) \quad
 \mathcal{I}_\rmi{b} & \!\!\equiv\!\! & 
 \Tint{PQ} \frac{1}{Q^2P^2(P-K)^2}
 \;,
 \\
 \TopoST(\Lqu,\Aqu) \times
 \TopoSB(\Lqu,\Asc,\Aqu) \quad
 \widetilde{\mathcal{I}}_\rmi{b} & \!\!\equiv\!\! & 
 \Tint{P\{R\}} \frac{1}{R^2P^2(P-K)^2}
 \;,
 \\
 \TopoST(\Lqu,\Asc) \times
 \TopoSTd(\Lqu,\Asc) \quad
 \mathcal{I}_\rmi{c} & \!\!\equiv\!\! & 
 \Tint{PQ} \frac{1}{Q^2P^4}
 \;, 
 \\
 \TopoST(\Lqu,\Aqu) \times
 \TopoSTd(\Lqu,\Asc) \quad
 \widetilde{\mathcal{I}}_\rmi{c} & \!\!\equiv\!\! & 
 \Tint{P\{R\}} \frac{1}{R^2P^4}
 \;,
 \\
 \TopoST(\Lqu,\Asc) \times
 \TopoSTd(\Lqu,\Aqu) \quad
 \widehat{\mathcal{I}}_\rmi{c} & \!\!\equiv\!\! & 
 \Tint{P\{R\}} \frac{1}{P^2R^4}
 \;, 
 \\
 \TopoST(\Lqu,\Aqu) \times
 \TopoSTd(\Lqu,\Aqu) \quad
 \overline{\mathcal{I}}_\rmi{c} & \!\!\equiv\!\! & 
 \Tint{\{RS\}} \frac{1}{S^2R^4}
 \;, 
 \\
 \TopoST(\Lqu,\Asc) \times
 \TopoSBd(\Lqu,\Asc,\Aqu) \quad
 \mathcal{I}_\rmi{d} & \!\!\equiv\!\! & 
 \Tint{PQ} \frac{K^2}{Q^2P^4(P-K)^2}
 \;,
 \\
 \TopoST(\Lqu,\Aqu) \times
 \TopoSBd(\Lqu,\Asc,\Aqu) \quad
 \widetilde{\mathcal{I}}_\rmi{d} & \!\!\equiv\!\! & 
 \Tint{P\{R\}} \frac{K^2}{R^2P^4(P-K)^2}
 \;,
 \\
 \TopoST(\Lqu,\Asc) \times
 \TopoSBd(\Lqu,\Aqu,\Asc) \quad
 \widehat{\mathcal{I}}_\rmi{d} & \!\!\equiv\!\! & 
 \Tint{P\{R\}} \frac{K^2}{P^2R^4(R-K)^2}
 \;,
 \\
 \TopoST(\Lqu,\Aqu) \times
 \TopoSBd(\Lqu,\Aqu,\Asc) \quad
 \overline{\mathcal{I}}_\rmi{d} & \!\!\equiv\!\! & 
 \Tint{\{RS\}} \frac{K^2}{S^2R^4(R-K)^2}
 \;,
 \\
 \ToptVS(\Asc,\Asc,\Lsc) \quad\;
 \mathcal{I}_\rmi{e} & \!\!\equiv\!\! & 
 \Tint{PQ} \frac{1}{Q^2P^2(P-Q)^2}
 \;,
 \\
 \ToptVS(\Aqu,\Aqu,\Lsc) \quad\;
 \widetilde{\mathcal{I}}_\rmi{e} & \!\!\equiv\!\! & 
 \Tint{P\{R\}} \frac{1}{R^2P^2(P-R)^2}
 \\[2mm] 
 \ToptSS(\Lqu,\Asc,\Asc,\Lqu) \quad
 \mathcal{I}_\rmi{f} & \!\!\equiv\!\! & 
 \Tint{PQ} \frac{1}{Q^2(Q-P)^2(P-K)^2}
 \;, 
 \\
 \ToptSS(\Lqu,\Aqu,\Aqu,\Lqu) \quad
 \widetilde{\mathcal{I}}_\rmi{f}\, & \!\!\equiv\!\! & 
 \Tint{P\{R\}} \frac{1}{R^2(R-P)^2(P-K)^2}
 \;, 
 \\ 
%
%
% \ToptSE(\Lqu,\Asc,\Asc,\Aqu,\Aqu) \quad
 \TopoSB(\Lqu,\Asc,\Aqu) \times
 \TopoSB(\Lqu,\Asc,\Aqu) \quad
 \mathcal{I}_\rmi{g} & \!\!\equiv\!\! & 
 \Tint{PQ} \frac{K^2}{P^2(P-K)^2Q^2(Q-K)^2}
 \;,
 \\
 \ToptSAr(\Lqu,\Asc,\Asc,\Aqu,\Asc) \quad
 \mathcal{I}_\rmi{h} & \!\!\equiv\!\! & 
 \Tint{PQ} \frac{K^2}{Q^2P^2(Q-P)^2(P-K)^2}
 \;, 
 \\
 \ToptSAr(\Lqu,\Aqu,\Asc,\Aqu,\Aqu) \quad
 \widetilde{\mathcal{I}}_\rmi{h} & \!\!\equiv\!\! & 
 \Tint{P\{R\}} \frac{K^2}{R^2P^2(R-P)^2(P-K)^2}
 \;, 
 \\
 \ToptSAr(\Lqu,\Aaqu,\Aaqu,\Asc,\Asc) \quad
 \widehat{\mathcal{I}}_\rmi{h} & \!\!\equiv\!\! & 
 \Tint{P\{R\}} \frac{K^2}{R^2P^2(R-P)^2(R-K)^2}
 \;, 
 \\ 
 \ToptSArx(\Lqu,\Aaqu,\Aaqu,\Asc,\Asc) \quad
 \widehat{\mathcal{I}}_\rmi{h'} \! & \!\!\equiv\!\! & 
 \Tint{P\{R\}} \frac{2K\cdot P}{R^2P^2(R-P)^2(R-K)^2}
 \;, 
 \\
 \ToptSM(\Lqu,\Asc,\Asc,\Aqu,\Aqu,\Lsc) \quad
 \mathcal{I}_\rmi{j} & \!\!\equiv\!\! & 
 \Tint{PQ} \frac{K^4}{Q^2P^2(Q-P)^2(P-K)^2(Q-K)^2} 
 \;.  
\ea

%%%%%%%%%%%%%%%%%%%%%%%%%%%%%% SECTION %%%%%%%%%%%%%%%%%%%%%%%%%%%%%%%%%
%
\section{Results for master spectral functions}
\la{app:B}

The spectral functions corresponding to the
structures of appendix~A are obtained from 
\be
 \rho^{ }_{\mathcal{I}_\rmi{x}} \equiv
 \im [ \mathcal{I}_\rmi{x} ]^{ }_{k^{ }_n \to -i [k^0 + i 0^+]} 
 \;. 
\ee
As has been elaborated upon in \se\ref{se:method}, they can 
be derived by carrying out the Matsubara sums, which corresponds
to cutting zero, one, or two lines and weighting them with a thermal
distribution. In the result, cuts correspond to logarithmic
terms, which can only arise from vacuum loops. Given that 2-cut 
contributions have no vacuum loop left, spectral functions
only arise from 0-cut and 1-cut contributions. 
After an expansion in small thermal momenta, 
the vacuum
parts always have simple cuts, and thereby the spectral functions
corresponding to the master structures of appendix~A are 
elementary, cf.\ \eqs\nr{rhoJa}--\nr{rhoIj} below.  

Before presenting the list, we would like to mention that apart
from the procedure described in \se\ref{se:method}, there exists 
also a more elaborate possibility for determining the spectral 
functions. Namely, one can first ``blindly'' work out the OPE expansion 
in the Euclidean domain up to the desired order in $1/K^2$. 
The corresponding results are listed in appendix~C. In these
results, non-analytic $K$-dependence only appears in a factor $X$, 
defined in \eq\nr{defXY}. Introducing the $\msbar$ scale parameter, 
it can be expanded as 
\be
 X = \frac{\mu^{-2\epsilon}}{(4\pi)^2} 
 \biggl( \frac{\bmu^2}{K^2} \biggr)^\epsilon
 \biggl[ 1- \frac{\pi^2\epsilon^2}{12} + \rmO(\epsilon^3) \biggr]
 \;,
\ee
and the corresponding spectral function then arises from 
\be
 \im[X]_{k_n \to -i [k^0+i0^+]}
 =  \sign(k^0) \, \frac{\mu^{-2\epsilon}}{16\pi}
 \biggl[\epsilon + \epsilon^2 \ln \frac{\bmu^2}{\mathcal{K}^2}
  + \rmO(\epsilon^3)
 \biggr] 
 \;. \la{mom_cut}
\ee
So, we observe that spectral functions can only arise from 
structures of type $X^n/\epsilon^m$, and that
all Euclidean structures that are either finite or have 
$1/\epsilon$-poles with the non-analytic scale dependence 
given by $T$ rather than by $K$, 
yield vanishing spectral functions. 

Proceeding to the list, we wish to remove 
clutter by not showing 
the non-consequential factor $\mu^{-2\epsilon}$
explicitly, and by similarly omitting 
the arguments of the functions $\nB{}$, $\nF{}$. 
In addition, the errors of the 2-loop structures, 
which are $\rmO(\epsilon,\frac{1}{\mathcal{K}^4})$, are 
not displayed.  
Thereby the spectral functions corresponding 
to the master structures read
\ba
% \TopoST(\Lqu,\Asc) \quad
 \rho^{ }_{\mathcal{J}_\rmi{a}} & \!\! = \!\! & 
% 0
% \;,
% \\
% 
% \TopoST(\Lqu,\Aqu) \quad
 \rho^{ }_{\widetilde{\mathcal{J}}_\rmi{a}}
% & \!\! = \!\! & 
 \; = \;
 0
 \;, \la{rhoJa}
 \\ 
%
% \TopoSB(\Lqu,\Asc,\Aqu) \quad
 \rho^{ }_{\mathcal{J}_\rmi{b}} & \!\! = \!\! & 
 -\frac{\mufac\mathcal{K}^2}{16\pi} \biggl[ 
   1 + \epsilon\biggl( \ln\frac{\bmu^2}{\mathcal{K}^2} + 2 \biggr)
 \biggr] 
 + \rmO\Bigl( \epsilon^2, e^{-\frac{k^0}{T}} \Bigr)
 \;,
 \\
%
% \TopoST(\Lqu,\Asc) \times
% \TopoSB(\Lqu,\Asc,\Aqu) \quad
 \rho^{ }_{\mathcal{I}_\rmi{b}} & \!\! = \!\! & 
  \int_\vec{p} \frac{\nB{}}{16\pi p}
 \error
 \;,
 \\
%
% \TopoST(\Lqu,\Aqu) \times
% \TopoSB(\Lqu,\Asc,\Aqu) \quad
 \rho^{ }_{\widetilde{\mathcal{I}}_\rmi{b}} & \!\! = \!\! & 
 -  \int_\vec{p} \frac{\nF{}}{16\pi p}
 \error
 \;,
 \\
%
% \TopoST(\Lqu,\Asc) \times
% \TopoSTd(\Lqu,\Asc) \quad
 \rho^{ }_{\mathcal{I}_\rmi{c}} & \!\! = \!\! & 
% 0
% \;, 
% \\
%
% \TopoST(\Lqu,\Aqu) \times
% \TopoSTd(\Lqu,\Asc) \quad
 \rho^{ }_{\widetilde{\mathcal{I}}_\rmi{c}}
% & \!\! = \!\! & 
% 0
% \;,
% \\
 \; = \; 
%
% \TopoST(\Lqu,\Asc) \times
% \TopoSTd(\Lqu,\Aqu) \quad
 \rho^{ }_{\widehat{\mathcal{I}}_\rmi{c}}
% & \!\! = \!\! & 
% 0
% \;, 
% \\
 \; = \; 
%
% \TopoST(\Lqu,\Aqu) \times
% \TopoSTd(\Lqu,\Aqu) \quad
 \rho^{ }_{\overline{\mathcal{I}}_\rmi{c}}
 % & \!\! = \!\! & 
 \; = \; 
 0
 \;, 
 \\
%
% \TopoST(\Lqu,\Asc) \times
% \TopoSBd(\Lqu,\Asc,\Aqu) \quad
 \rho^{ }_{\mathcal{I}_\rmi{d}} & \!\! = \!\! & 
 -  \int_\vec{p} \frac{\nB{}}{16\pi p}
 \error
 \;,
 \\
% 
% \TopoST(\Lqu,\Aqu) \times
% \TopoSBd(\Lqu,\Asc,\Aqu) \quad
 \rho^{ }_{\widetilde{\mathcal{I}}_\rmi{d}} & \!\! = \!\! & 
   \int_\vec{p} \frac{\nF{}}{16\pi p}
 \error
 \;,
 \\
%
% \TopoST(\Lqu,\Asc) \times
% \TopoSBd(\Lqu,\Aqu,\Asc) \quad
 \rho^{ }_{\widehat{\mathcal{I}}_\rmi{d}} & \!\! = \!\! & 
 -  \int_\vec{p} \frac{\nB{}}{16\pi p}
 \error
 \;,
 \\
%
% \TopoST(\Lqu,\Aqu) \times
% \TopoSBd(\Lqu,\Aqu,\Asc) \quad
 \rho^{ }_{\overline{\mathcal{I}}_\rmi{d}} & \!\! = \!\! & 
   \int_\vec{p} \frac{\nF{}}{16\pi p}
 \error
 \;,
 \\
%
% \ToptVS(\Asc,\Asc,\Lsc) \quad\;
 \rho^{ }_{\mathcal{I}_\rmi{e}}  & \!\! = \!\! & 
 % 0
 % \;,
 % \\
% 
% \ToptVS(\Aqu,\Aqu,\Lsc) \quad\;
 \rho^{ }_{\widetilde{\mathcal{I}}_\rmi{e}}
 % & \!\! = \!\! & 
 \; = \;
 0 
 \;, 
 \\[2mm] 
%
% \ToptSS(\Lqu,\Asc,\Asc,\Lqu) \quad
 \rho^{ }_{\mathcal{I}_\rmi{f}} & \!\! = \!\! & 
 \frac{\mufac\mathcal{K}^2}{8(4\pi)^3}
 + 
 \int_\vec{p} \frac{2\nB{} - \nF{}}{16\pi p}
 \error
 \;, 
 \\
% 
% \ToptSS(\Lqu,\Aqu,\Aqu,\Lqu) \quad
 \rho^{ }_{\widetilde{\mathcal{I}}_\rmi{f}}\, & \!\! = \!\! & 
 \frac{\mufac\mathcal{K}^2}{8(4\pi)^3}
 - 
  \int_\vec{p} \frac{3 \nF{}}{16\pi p}
 \error
 \;, 
 \\ 
% 
% \TopoSB(\Lqu,\Asc,\Aqu) \times
% \TopoSB(\Lqu,\Asc,\Aqu) \quad
 \rho^{ }_{\mathcal{I}_\rmi{g}} & \!\! = \!\! & 
  -\frac{\mufac\mathcal{K}^2}{2(4\pi)^3} \biggl( 
   \frac{1}{\epsilon}+ 2\ln\frac{\bmu^2}{\mathcal{K}^2} + 4 \biggr)
  +   
  \int_\vec{p} \frac{\nB{} - \nF{}}{8\pi p}
  + \frac{k_0^2 + k^2/3}{2\pi \mathcal{K}^4}
 \int_\vec{p} p\, (\nB{}-\nF{})
 \error
 \;,
 \\
%
% \ToptSAr(\Lqu,\Asc,\Asc,\Aqu,\Asc) \quad
 \rho^{ }_{\mathcal{I}_\rmi{h}} & \!\! = \!\! & 
  -\frac{\mufac\mathcal{K}^2}{4(4\pi)^3} \biggl( 
   \frac{1}{\epsilon}+ 2\ln\frac{\bmu^2}{\mathcal{K}^2} + 5 \biggr)
  -  
  \int_\vec{p} \frac{2 \nB{} + \nF{}}{16\pi p}
  -  \frac{k_0^2 + k^2/3}{2\pi \mathcal{K}^4}
 \int_\vec{p} p\, \Bigl(\frac{\nB{}}{3}+\frac{\nF{}}{2}\Bigr)
 \error
 \;, \hspace*{7mm}
 \\
%
% \ToptSAr(\Lqu,\Aqu,\Asc,\Aqu,\Aqu) \quad
 \rho^{ }_{\widetilde{\mathcal{I}}_\rmi{h}} & \!\! = \!\! & 
  -\frac{\mufac\mathcal{K}^2}{4(4\pi)^3} \biggl( 
   \frac{1}{\epsilon}+ 2\ln\frac{\bmu^2}{\mathcal{K}^2} + 5 \biggr)
  +  
  \int_\vec{p} \frac{\nF{}}{16\pi p}
  -  \frac{k_0^2 + k^2/3}{2\pi \mathcal{K}^4}
 \int_\vec{p} p\, \Bigl(\frac{\nF{}}{6}\Bigr)
 \error
 \;, 
 \\
%
% \ToptSAr(\Lqu,\Aaqu,\Aaqu,\Asc,\Asc) \quad
 \rho^{ }_{\widehat{\mathcal{I}}_\rmi{h}} & \!\! = \!\! & 
  -\frac{\mufac\mathcal{K}^2}{4(4\pi)^3} \biggl( 
   \frac{1}{\epsilon}+ 2\ln\frac{\bmu^2}{\mathcal{K}^2} + 5 \biggr)
  +   
  \int_\vec{p} \frac{\nF{}}{16\pi p}
  +  \frac{k_0^2 + k^2/3}{2\pi \mathcal{K}^4}
 \int_\vec{p} p\, \Bigl(\frac{\nB{}}{3}+\frac{\nF{}}{6}\Bigr)
 \error
 \;, 
 \\ 
% 
% \ToptSArx(\Lqu,\Aaqu,\Aaqu,\Asc,\Asc) \quad
 \rho^{ }_{\widehat{\mathcal{I}}_\rmi{h'}} \! & \!\! = \!\! & 
  -\frac{\mufac\mathcal{K}^2}{8(4\pi)^3} \biggl( 
   \frac{1}{\epsilon}+ 2\ln\frac{\bmu^2}{\mathcal{K}^2} + \fr92 \biggr)
  +   
  \int_\vec{p} \frac{\nB{}}{16\pi p}
  -  \frac{k_0^2 + k^2/3}{2\pi \mathcal{K}^4}
 \int_\vec{p} p\, \Bigl(\frac{\nF{}}{12}\Bigr)
 \error
 \;, 
 \\
%
% \ToptSM(\Lqu,\Asc,\Asc,\Aqu,\Aqu,\Lsc) \quad
 \rho^{ }_{\mathcal{I}_\rmi{j}} & \!\! = \!\! &  
  \int_\vec{p} \frac{\nF{}-2\nB{}}{8\pi p}
  +   \frac{k_0^2 + k^2/3}{2\pi \mathcal{K}^4}
 \int_\vec{p} p\, \Bigl(\frac{11\nF{}}{6}-\frac{7\nB{}}{3}\Bigr)
 \error
 \;. \la{rhoIj}  
\ea

\mbox{ }

%%%%%%%%%%%%%%%%%%%%%%%%%%%%%% SECTION %%%%%%%%%%%%%%%%%%%%%%%%%%%%%%%%%
%
\section{Euclidean large-momentum expansions}
\la{app:C}

For completeness, we list here Euclidean large-momentum expansions 
for the master sum-integrals defined in appendix A. In order to be
as concise as possible, we introduce the notation
\be
  n_i \equiv \left\{ \begin{array}{rl}
   \nB{} \;, & \mbox{bosonic line } \\ 
  -\nF{} \;, & \mbox{fermionic line } \end{array} \right.
 \;, 
\ee
and provide expressions valid simultaneously for all 
the statistics carried by the lines. The shorthands 
\be
  X \equiv 
 \frac{K^{-2\epsilon}}{(4\pi)^{2-\epsilon}}
 \frac{\Gamma(1+\epsilon) \Gamma^2(1-\epsilon)}{\Gamma(1-2\epsilon)}
  \;, \quad 
  Y \equiv
  \frac{\Gamma(1+2\epsilon)\Gamma^2(1-2\epsilon)}
  {\Gamma(1-3\epsilon)\Gamma^2(1+\epsilon)\Gamma(1-\epsilon)}
  \;, \la{defXY}
\ee
as well as 
\be
  \Theta_K \equiv 
  \frac{\frac{k^2}{3-2\epsilon} - k_n^2}{K^2}
\ee
are also helpful. Thereby we obtain
\ba
 \TopoSTu(\Lqu,\Aqq,1) 
 & = & 
 \int_\vec{p} \frac{n_1(p)}{p}
 \;,
 \\
 \TopoSBu(\Lqu,\Aqq,\Aqq,1,2)
 & = & 
 \frac{K^2 X}{\epsilon(1-2\epsilon)}
 + \sum_{i=1}^{2} \int_\vec{p} 
 \biggl[
   \frac{n_i(p)}{p} + \frac{4 \,\Theta_K\, p\, n_i(p)}{K^2}   
 \biggr] + \rmO\Bigl( \frac{T^6}{K^4} \Bigr)
 \;,
 \\
 \TopoSTu(\Lqu,\Aqq,3) \times
 \TopoSBu(\Lqu,\Aqq,\Aqq,1,2)
 & = & 
 \frac{X}{\epsilon(1-2\epsilon)} \int_\vec{p} \frac{n_3(p)}{p}
 + \frac{1}{K^2} \sum_{i=1}^{2} \int_\vec{p,q}
   \frac{n_3(p) n_i(q) }{pq}
 + \rmO\Bigl( \frac{T^6}{K^4} \Bigr)
 \;, \hspace*{1cm}
 \\
 \TopoSTu(\Lqu,\Aqq,1) \times
 \TopoSTdu(\Lqu,\Aqq,2)
 & = & 
 \frac{1-2\epsilon}{2}
  \int_\vec{p,q}
  \frac{n_1(p) n_2(q) }{pq^3}
 \;, 
 \\
 \TopoSTu(\Lqu,\Aqq,3) \times
 \TopoSBdu(\Lqu,\Aqq,\Aqq,1,2)
 & = & 
 -\frac{X}{\epsilon} \int_\vec{p} \frac{n_3(p)}{p}
 +\frac{1-2\epsilon}{2}
  \int_\vec{p,q}
  \frac{n_3(p) n_1(q) }{pq^3}
 \nn 
 & + & 
  \frac{
    \epsilon + 2 (1-\epsilon)(3-2\epsilon) \,\Theta_K 
   }{(2-\epsilon)K^2}
  \int_\vec{p,q}
  \frac{n_3(p) n_1(q) }{pq}
 \nn
 & + &  
  \frac{1}{K^2} 
  \int_\vec{p,q}
  \frac{n_3(p) n_2(q) }{pq}
 + \rmO\Bigl( \frac{T^6}{K^4} \Bigr)
 \;,
 \\
 \ToptVS(\Asc,\Asc,\Lsc) \;  = \;  
 \ToptVS(\Aqu,\Aqu,\Lsc)
 & = & 
 0
 \\[2mm] 
 \ToptSSu(\Lqu,\Aqq,\Aqq,\Lqq,1,2,3)
 & = & 
 - \frac{K^2 X^2 Y}{2\epsilon(1-2\epsilon)(1-3\epsilon)(2-3\epsilon)}
 \nn & + &  
 \frac{X}{\epsilon(1-2\epsilon)} \sum_{i=1}^3 
 \int_\vec{p} \biggl[ 
   \frac{n_i(p)}{p} +
   \frac{2\epsilon (1+\epsilon)\,\Theta_K\, p \, n_i(p)}{K^2} 
 \biggr]
 \nn & + & 
 \frac{1}{K^2} \sum_{i > j} 
  \int_\vec{p,q}
  \frac{n_i(p) n_j(q) }{pq}
 + \rmO\Bigl( \frac{T^6}{K^4} \Bigr)
 \;, 
 \\
 \TopoSBu(\Lqu,\Aqq,\Aqq,1,2) \times
 \TopoSBu(\Lqu,\Aqq,\Aqq,3,4) 
 & = & 
 \frac{K^2 X^2}{\epsilon^2(1-2\epsilon)^2}
 \nn & + &  
 \frac{X}{\epsilon(1-2\epsilon)} \sum_{i=1}^4 
 \int_\vec{p} \biggl[ 
   \frac{n_i(p)}{p} +
   \frac{4 \,\Theta_K\, p \, n_i(p)}{K^2} 
 \biggr]
 \nn & + &  
 \frac{1}{K^2} 
  \int_\vec{p,q}
  \frac{(n_1+n_2)(p)\, (n_3+n_4)(q) }{pq}
 + \rmO\Bigl( \frac{T^6}{K^4} \Bigr)
 \;,
 \\
 \ToptSAru(\Lqu,\Aqq,\Aqq,\Aqq,\Aqq,4,2,1,3)
 & = & 
 \frac{K^2 X^2 Y}{2\epsilon^2(1-2\epsilon)(1-3\epsilon)}
 \nn & +  & 
 \frac{X}{\epsilon (1-2\epsilon)} 
 \int_\vec{p} \biggl[ 
   \frac{n_1(p)}{p} +
   \frac{2(1+\epsilon)(2+\epsilon)\,\Theta_K\, p \, n_1(p)}{K^2} 
 \biggr]
 \nn & - &  
 \frac{X}{\epsilon} \sum_{i=3}^4 
 \int_\vec{p} \biggl[ 
   \frac{n_i(p)}{p} +
   \frac{2(1+\epsilon)(2+\epsilon)\,\Theta_K\, p \, n_i(p)}{3 K^2} 
 \biggr]
 \nn & + &     
 \frac{\epsilon^2 + 4(1-\epsilon)^2 \,\Theta_K}{\epsilon(2-\epsilon)K^2} 
  \int_\vec{p,q}
  \frac{n_3(p) n_4(q) }{pq}
 \nn & + & 
 \frac{1}{K^2} \sum_{i =3}^{4} 
  \int_\vec{p,q}
  \frac{n_1(p) n_i(q) }{pq}
 + \rmO\Bigl( \frac{T^6}{K^4} \Bigr)
 \;, 
 \\
 \ToptSArxu(\Lqu,\Aqq,\Aqq,\Aqq,\Aqq,4,2,1,3)
 & = & 
 \frac{K^2 X^2 Y}{2\epsilon^2(1-3\epsilon)(2-3\epsilon)}
 \nn & +  & 
 \frac{X}{\epsilon (1-2\epsilon)} 
 \int_\vec{p} \biggl[ 
   \frac{n_1(p)}{p} +
   \frac{2(1+\epsilon)^2\,\Theta_K\, p \, n_1(p)}{K^2} 
 \biggr]
 \nn & - &
 \frac{X}{\epsilon} 
 \int_\vec{p}
 \frac{2 (1+\epsilon) \,\Theta_K\, p \, n_3(p)}{K^2} 
 \nn & + &  
 \frac{2 X}{1-2\epsilon} 
 \int_\vec{p} \biggl[ 
   \frac{n_4(p)}{p} +
   \frac{(1+\epsilon)(1+2\epsilon^2)\,\Theta_K\, p \, n_4(p)}{3 \epsilon K^2} 
 \biggr]
 \nn & + &     
 \frac{\epsilon + 2(1-\epsilon)^2 \,\Theta_K}{\epsilon(2-\epsilon)K^2} 
  \int_\vec{p,q}
  \frac{(n_2+n_3)(p) n_4(q) - n_2(p)n_3(q) }{pq}
 \nn & + & 
   \frac{2}{K^2} 
  \int_\vec{p,q}
  \frac{n_1(p) n_4 (q) }{pq}
 + \rmO\Bigl( \frac{T^6}{K^4} \Bigr)
 \;, 
 \\
 \ToptSMu(\Lqu,\Aqq,\Aqq,\Aqq,\Aqq,\Lqq) 
 & = & 
 \frac{K^2 X^2(1- Y)}{\epsilon^3(1-2\epsilon)}
 \nn & -  & 
 \frac{X}{\epsilon} \sum_{i=1}^{4}
 \int_\vec{p} \biggl[ 
   \frac{n_i(p)}{p} +
   \frac{2(11+6\epsilon+\epsilon^2)\,\Theta_K\, p \, n_i(p)}{3 K^2} 
 \biggr]
 \nn & - &  
 \frac{2 X (1+\epsilon)}{\epsilon} 
 \int_\vec{p} \biggl[ 
   \frac{n_5(p)}{p} +
   \frac{(2+\epsilon)(3+\epsilon)\,\Theta_K\, p \, n_5(p)}{3 K^2} 
 \biggr]
 \nn & - &
 \frac{\epsilon + 2(1-\epsilon)^2 \,\Theta_K}{\epsilon(2-\epsilon)K^2} 
  \int_\vec{p,q}
  \frac{n_1(p) n_2(q) + n_3(p)n_4(q) }{pq}
 \nn & + & 
 \frac{\epsilon(1+\epsilon)
 + 6(1-\epsilon)^2 \,\Theta_K}{\epsilon(2-\epsilon)K^2} \sum_{i=1}^{4}
  \int_\vec{p,q}
  \frac{n_i(p)n_5(q) }{pq}
 \nn & + & 
   \frac{1}{K^2} 
  \int_\vec{p,q}
  \frac{n_1(p) n_3 (q) +n_2(p) n_4 (q) }{pq}
 + \rmO\Bigl( \frac{T^6}{K^4} \Bigr)
 \;.  
\ea

%%%%%%%%%%%%%%%%%%%%%%%%%%%%%% SECTION %%%%%%%%%%%%%%%%%%%%%%%%%%%%%%%%%
%
\section{On the treatment of Dirac traces}
\la{app:D}

In analogy with the leading-order example of \eq\nr{g5PiE_0}, 
let us compare the NDR expressions in \eqs\nr{EleB}--\nr{EleG}
with the recipe explained below \eq\nr{g5Larin} as well as
with the strict 't~Hooft - Veltman scheme.
After Wick contractions, Lorentz algebra, isospin traces 
and using standard properties of non-chiral Dirac matrices
(but doing nothing with $\gamma_5$ for the moment), 
the contributions to the Euclidean correlator $\Pi_E(K)$ 
in Feynman gauge read (omitting the overall factor 
$|h_{\nu\rmii{B}}|^2$, and abbreviating Lorentz-indices\footnote{%
Note that we do not use $\mu_5$ here, to avoid confusion of 
$\ga_{\mu_5}$ with $\ga_5$.
 } $\mu_1,...,\mu_6$
as in $\ga_1\equiv\ga_{\mu_1}$ or $K_2\equiv K_{\mu_2}$ etc.)
\ba
%\EleA &=& -2 
%%|h_{\nu\rmii{B}}|^2\,
%  \tr[\ga_1\aL \ga_2\aR ]\,
%  \Tint{P} \frac{K_1(K-P)_2}{P^2(K-P)^2}\;, \\
\EleB &=& 12 
%|h_{\nu\rmii{B}}|^2 
  \lambda_\rmii{B}\,
  \tr[\ga_1\aL\! \ga_2\,\aR ]\,
  \Tint{PQ} \frac{K_1\,(K-P)_2}{Q^2P^4(K-P)^2}\;, \la{d1} \\
\EleC &=& 2 
%|h_{\nu\rmii{B}}|^2 
  |h_{t\rmii{B}}|^2 \Nc\,
  \tr[\ga_1\aL\!\ga_2\,\aR ]\,\tr[\ga_3\aL\!\ga_4\,\aR ]\,
  \Tint{P\{R\}} \frac{K_1(K-P)_2R_3(P-R)_4}{P^4(K-P)^2R^2(P-R)^2}\;,
 \la{closed_f}
 \hspace*{10mm} \\
\EleD &=& 
%|h_{\nu\rmii{B}}|^2 
  (g_{1\rmii{B}}^2 + 3 g_{2\rmii{B}}^2)\,
  \tr[\ga_1\aL\!\ga_2\,\aR ]\, \frac{D}{2}\,
  \Tint{PQ} \frac{K_1(K-P)_2}{Q^2P^4(K-P)^2}\;, \\
\EleE &=& -
%|h_{\nu\rmii{B}}|^2 
  (g_{1\rmii{B}}^2 + 3 g_{2\rmii{B}}^2)\,
  \tr[\ga_1\aL\!\ga_2\,\aR ]\, \frac12\,
  \Tint{PQ} \frac{K_1(K-P)_2(P+Q)^2}{P^4Q^2(P-Q)^2(K-P)^2}\;, \\
\EleF &=& 
%|h_{\nu\rmii{B}}|^2\,
  s_{1234}\, \frac12\, 
  \Tint{PQ} \frac{K_1(K-P)_2(K-Q)_3(K-P)_4}{P^2(P-Q)^2(K-P)^4(K-Q)^2}\;,
\\&& s_{1234}\;\equiv\; 
  (g_{1\rmii{B}}^2 + 3 g_{2\rmii{B}}^2)\,(D-2)\,
  \tr[\ga_1\aL\!\ga_6\,\aR ]
  \,(g_{23}g_{46}-g_{24}g_{36}+g_{34}g_{26})
\nonumber\\&&\hphantom{s_{1234}\;}
  -\;g_{1\rmii{B}}^2\,\tr[\ga_1\aL \ga_2\ga_6\,\aR \ga_3\ga_6\,\aR \ga_4\aR]
\nonumber\\&&\hphantom{s_{1234}\;}
  -\;3g_{2\rmii{B}}^2\tr[\ga_1\aL\! \ga_2\ga_6
  \,(\aL\! \ga_3\ga_6\aL -\ga_3\ga_6)\,\ga_4\aR]\,, 
  \la{s1234} \\
\EleG &=& 
%|h_{\nu\rmii{B}}|^2\,
  t_{1234}\, \frac12\,
  \Tint{PQ} \frac{K_1(K-Q)_2(P+Q)_3(K-P)_4}{P^2Q^2(P-Q)^2(K-P)^2(K-Q)^2}\;,
 \la{d7}
\\&& t_{1234}\;\equiv\; 
  (g_{1\rmii{B}}^2 + 3 g_{2\rmii{B}}^2)\,\tr[\ga_1\aL\!\ga_6\,\aR ]\,
  (g_{23}g_{46}-g_{24}g_{36}+g_{34}g_{26})
\nonumber\\&& \hphantom{t_{1234}\;}
  -\;3g_{2\rmii{B}}^2\,\tr[\ga_1\aL \ga_2\ga_3\aR \ga_4\aR ]\;. \la{t1234}
\ea
All diagrams with Higgs self-energy insertions are seen to be
proportional to the same structure that already appeared in 
the LO contribution; if handled as shown below \eq\nr{g5PiE_0},
they lead to $\tr[\ga_1\aL \ga_2\aR ]\rightarrow 2g_{12}$ 
and immediately reduce to the NDR results.
On the other hand, the two diagram classes with the vectors coupling 
to the lepton line do seem to get additional contributions proportional 
to different Dirac traces, which have been separated in the second 
and third lines
of \eqs\nr{s1234} and \nr{t1234}.

It turns out, however, that upon employing the prescription explained
below \eq\nr{g5Larin} these additional contributions vanish 
identically, which leaves us with the same results as in NDR
for each of the diagrams.
As a specific example, consider the second line of \eq\nr{t1234},
\ba\la{g5Ex}
8\,\tr[\ga_1\aL \ga_2\ga_3\aR \ga_4\aR ] 
&=&
 \tr\bigl[\ga_1\ga_2\ga_3(\ga_4+\ga_5\ga_4\ga_5)\bigr]
 -\tr\bigl[\ga_1\ga_5\ga_2\ga_3(\ga_4+\ga_5\ga_4\ga_5)\bigr]
\nonumber\\
 &-&\tr\bigl[\ga_1\ga_5\ga_2\ga_3(\ga_5\ga_4+\ga_4\ga_5)\bigr]
 +\tr\bigl[\ga_1\ga_2\ga_3(\ga_5\ga_4+\ga_4\ga_5)\bigr]\nonumber\\
&=& 
 \tr\bigl[\ga_4\ga_1\ga_2\ga_3\ga_5
 +\ga_1\ga_2\ga_3\ga_4\ga_5\bigr]\;,
\ea
where we have used an anticommuting $\ga_5$ as well as $\ga_5^2={1}$ 
in traces with more than one~$\ga_5$. 
Then, for the remaining traces with a single $\ga_5$ we get 
$4(\varepsilon_{4123}+\varepsilon_{1234})$, 
such that the two terms 
cancel due to antisymmetry of the $\varepsilon$ tensor.
In a completely analogous way, the last two lines of \eq\nr{s1234}
are seen not to contribute in this specific scheme.

If the 't Hooft - Veltman scheme is used rather than the 
recipe below \eq\nr{g5Larin}, then graphs with closed 
fermion loops {\em do differ} from those in NDR. To see this, 
note that the integral in \eq\nr{closed_f}  
evaluates at zero temperature to
\ba
 && \hspace*{-2cm}
 \int_{PR} \frac{K_1(K-P)_2R_3(P-R)_4}{P^4(K-P)^2R^2(P-R)^2} 
 =  
% \frac{(1-\epsilon) X^2 Y K_1}
% {24\epsilon^2(1+\epsilon)(1-2\epsilon)(1-3\epsilon)
%    (2-3\epsilon)(3-2\epsilon)} 
 \frac{X^2 Y K_1}{144}
 \biggl( \frac{1}{\epsilon^2} + \frac{31}{6\epsilon}
 + \frac{763}{36} + \rmO(\epsilon) \biggr)
 \nn 
 && \; \times \,  
 \biggl[
   4\epsilon(1-2\epsilon) \frac{K_2 K_3 K_4}{K^2}
   + (5+2\epsilon) K_2\, g_{34}
   - (1-2\epsilon) 
   \bigl( K_3\, g_{24} + K_4\, g_{23} \bigr)
 \biggr]
 \;,  \la{closed_f_2}
\ea
where the coefficients $X$ and $Y$ are defined in \eq\nr{defXY}. With 
$\tr[\ga_3\aL\!\ga_4\,\aR ] \to 2g_{34}$ this yields
the NDR result, but with \eq\nr{problem} we rather have  
$\tr[\ga_3\aL\!\ga_4 \aR ] = 2\tilde g_{34}$, 
where $\tilde g_{\mu\nu} \equiv 1$, 
if $\mu = \nu \le 3$, and $\tilde g_{\mu\nu} \equiv 0$ otherwise.  
Contracting with the $g_{34}$ in \eq\nr{closed_f_2} produces 
$4$ rather than $4-2\epsilon$, which turns into 
a difference of $\rmO(1)$ in the spectral function because
of the prefactor $1/\epsilon^2$ (concretely, $\fr72$ in 
\eq\nr{pre_result} turns into $\fr{73}{18}$). That said,  this 
difference can presumably be ``hidden'' if 
the neutrino Yukawa coupling is expressed in terms of a physical 
quantity (such as a pole mass) through a computation carried out 
in the same scheme. 

In thermal corrections to \eq\nr{closed_f_2}, 
there is only a prefactor $1/\epsilon$ in the Euclidean 
domain (cf.\ appendix~C), which implies that the corresponding
spectral function is finite (cf.\ appendix~B). 
Therefore we expect that the ambiguity of $\rmO(\epsilon)$ 
in a prefactor does not affect NLO thermal corrections
to spectral functions. Nevertheless, it might be interesting
to work out the full tensor integrals of \eqs\nr{d1}--\nr{d7} 
in the OPE regime, in analogy with the expansions in appendix~C, 
thereby producing results for any desirable scheme.  
Unfortunately this involves a substantial amount of work 
and goes beyond the scope of the present study. 

%%%%%%%%%%%%%%%%%%%%%%%%%%%%%%%%%%%%%%%%%%%%%%%%%%%%%%%%%%%%%%%%%%%%%%%%%%%
%


\begin{thebibliography}{99}

\bibitem{rev1}
  F.D.~Steffen,
  {\em Dark Matter Candidates -- Axions, Neutralinos,  
  Gravitinos, and Axinos,}
  Eur.\ Phys.\ J.\  {C 59} (2009)  557
  [0811.3347].
  %%CITATION = EPHJA,C59,557;%%


\bibitem{rev2}
  A.~Boyarsky, O.~Ruchayskiy and M.~Shaposhnikov,
  {\em The Role of Sterile Neutrinos in Cosmology and Astrophysics,}
  Ann.\ Rev.\ Nucl.\ Part.\ Sci.\  {59} (2009) 191
  [0901.0011].
  %%CITATION = ARNUA,59,191;%%

\bibitem{yanagida}
  M.~Fukugita and T.~Yanagida,
  {\em Baryogenesis Without Grand Unification,}
  Phys.\ Lett.\  B {174} (1986) 45.
  %%CITATION = PHLTA,B174,45;%%


\bibitem{lepto1}
  W.~Buchm\"uller, R.D.~Peccei and T.~Yanagida,
  {\em Leptogenesis as the origin of matter,}
  Ann.\ Rev.\ Nucl.\ Part.\ Sci.\  {55 } (2005) 311
  [hep-ph/0502169].
  %%CITATION = ARNUA,55,311;%%

\bibitem{lepto2}
  S.~Davidson, E.~Nardi and Y.~Nir,
  {\em Leptogenesis,}
  Phys.\ Rept.\  {466 } (2008)  105
  [0802.2962].
  %%CITATION = PRPLC,466,105;%%

\bibitem{anisimov}
  A.~Anisimov, D.~Besak and D.~B\"odeker,
  {\it Thermal production of relativistic Majorana neutrinos: 
  Strong enhancement by multiple soft scattering,}
  JCAP {03} (2011) 042
  [1012.3784].
  %%CITATION = JCAPA,1103,042;%%

\bibitem{salvio}
  A.~Salvio, P.~Lodone and A.~Strumia,
  {\it Towards leptogenesis at NLO: 
  the right-handed neutrino interaction rate,}
  JHEP {08} (2011) 116
  [1106.2814].
  %%CITATION = JHEPA,1108,116;%%

\bibitem{cp2}
  M.~Garny, A.~Hohenegger and A.~Kartavtsev,
  {\em Medium corrections to the CP-violating parameter in leptogenesis,}
  Phys.\ Rev.\  {D 81} (2010)  085028
  [1002.0331].
  %%CITATION = PHRVA,D81,085028;%%

\bibitem{cp3}
  M.~Beneke, B.~Garbrecht, C.~Fidler, M.~Herranen and P.~Schwaller,
  {\em Flavoured Leptogenesis in the CTP Formalism,}
  Nucl.\ Phys.\  {B 843} (2011)  177
  [1007.4783].
  %%CITATION = NUPHA,B843,177;%%

\bibitem{cp4}
  C.S.~Fong, M.C.~Gonzalez-Garcia and J.~Racker,
  {\em CP Violation from Scatterings with Gauge Bosons in Leptogenesis,}
  Phys.\ Lett.\  {B 697} (2011)  463
  [1010.2209].
  %%CITATION = PHLTA,B697,463;%%

\bibitem{cp0}
  J.-S.~Gagnon and M.~Shaposhnikov,
  {\em Baryon Asymmetry of the Universe without
  Boltzmann or Kadanoff-Baym equations,}
  Phys.\ Rev.\  {D 83} (2011)  065021
  [1012.1126].
  %%CITATION = PHRVA,D83,065021;%%

\bibitem{cp1}
  A.~Anisimov, W.~Buchm\"uller, M.~Drewes and S.~Mendizabal,
  {\em Quantum Leptogenesis I,}
  Annals Phys.\  {326 } (2011)  1998
  [1012.5821].
  %%CITATION = APNYA,326,1998;%%

\bibitem{cpn}
  C.~Kiessig and M.~Pl\"umacher,
  {\em Hard-Thermal-Loop Corrections in Leptogenesis I: CP-Asymmetries,}  
  1111.1231.
  %%CITATION = ARXIV:1111.1231;%%

\bibitem{OPE}
  K.G.~Wilson and W.~Zimmermann,
  {\it Operator Product Expansions and Composite Field Operators
  in the General Framework of Quantum Field Theory,}
  Commun.\ Math.\ Phys.\  {24} (1972) 87.
  %%CITATION = CMPHA,24,87;%%

\bibitem{simon}
  S.~Caron-Huot,
  {\it Asymptotics of thermal spectral functions,}
  Phys.\ Rev.\  D {79} (2009) 125009
  [0903.3958].
  %%CITATION = PHRVA,D79,125009;%%

\bibitem{Bulk_OPE}
  M.~Laine, M.~Veps\"al\"ainen and A.~Vuorinen,
  {\it Ultraviolet asymptotics of scalar and pseudoscalar
  correlators in hot Yang-Mills theory,}
  JHEP {10} (2010) 010
  [1008.3263].
  %%CITATION = JHEPA,1010,010;%%

\bibitem{Shear_OPE}
  Y.~Schr\"oder, M.~Veps\"al\"ainen, A.~Vuorinen and Y.~Zhu,
  {\it The ultraviolet limit and sum rule for the shear correlator
  in hot Yang-Mills theory,}
  JHEP {12} (2011) 035
  [1109.6548].
  %%CITATION = ARXIV:1109.6548;%%

\bibitem{hadronic}
  T.~Asaka, M.~Laine and M.~Shaposhnikov,
  {\em On the hadronic contribution to sterile neutrino production,}
  JHEP {06} (2006) 053
  [hep-ph/0605209].
  %%CITATION = JHEPA,0606,053;%%

\bibitem{shifuller}
  M.~Laine and M.~Shaposhnikov,
  {\it Sterile neutrino dark matter as a consequence of 
  $\nu$MSM-induced lepton asymmetry,}
  JCAP {06} (2008) 031
  [0804.4543].
  %%CITATION = JCAPA,0806,031;%%

 \bibitem{lebellac}
  M. Le Bellac, {\it Thermal Field Theory} 
  (Cambridge University Press, Cambridge, 2000).
  %%CITATION = NONE;%%


\bibitem{tHV}
  G.~'t Hooft and M.J.G.~Veltman,
  {\em Regularization and Renormalization of Gauge Fields,}
  Nucl.\ Phys.\  {B 44} (1972)  189.
  %%CITATION = NUPHA,B44,189;%%
  
\bibitem{bm}
  P.~Breitenlohner and D.~Maison,
  {\em Dimensional Renormalization and the Action Principle,}
  Commun.\ Math.\ Phys.\ \ {52} (1977) 11.
  %%CITATION = CMPHA,52,11;%%

\bibitem{JK}
   J.G.~K\"orner, N.~Nasrallah and K.~Schilcher,
   {\em Evaluation of the flavor-changing vertex 
   $b \to s H$ using the Breitenlohner -- Maison -- 't Hooft -- Veltman 
   $\gamma_5$ scheme,}
   Phys.\ Rev.\  {D 41} (1990)  888.
   %%CITATION = PHRVA,D41,888;%%

\bibitem{BW}
  A.J.~Buras and P.H.~Weisz,
  {\em QCD Nonleading Corrections to Weak Decays in 
  Dimensional Regularization and 't Hooft-Veltman Schemes,}
  Nucl.\ Phys.\  {B 333} (1990)  66.
  %%CITATION = NUPHA,B333,66;%%

\bibitem{SL}
  S.A.~Larin,
  {\em The Renormalization of the axial anomaly
  in dimensional regularization,}
  Phys.\ Lett.\  {B 303} (1993)  113
  [hep-ph/9302240].
  %%CITATION = PHLTA,B303,113;%%

\bibitem{SLns}
  S.A.~Larin and J.A.M.~Vermaseren,
  {\em The $\alpha_s^3$ corrections to the Bjorken sum rule for polarized 
  electroproduction and to the Gross-Llewellyn Smith sum rule,}
  Phys.\ Lett.\ {B 259} (1991) 345.
  %%CITATION = PHLTA,B259,345;%%

\bibitem{ae}
  P.B.~Arnold and O.~Espinosa,
  {\em The Effective potential and first order phase transitions: 
  Beyond leading-order,}
  Phys.\ Rev.\ D\ {47} (1993) 3546
  [Erratum-ibid.\ D\ {50} (1994) 6662]
  [hep-ph/9212235].
  %%CITATION = PHRVA,D47,3546;%%

\bibitem{Bulk_wdep}
  M.~Laine, A.~Vuorinen and Y.~Zhu,
  {\em Next-to-leading order thermal spectral functions
  in the perturbative domain,}
  JHEP {09} (2011)  084
  [1108.1259].
  %%CITATION = JHEPA,1109,084;%%

\bibitem{pert}
  K.~Farakos, K.~Kajantie, K.~Rummukainen and M.E.~Shaposhnikov,
  {\it 3D physics and the electroweak phase transition: Perturbation theory,}
  Nucl.\ Phys.\ B\ {425} (1994) 67
  [hep-ph/9404201].
  %%CITATION = NUPHA,B425,67;%%

\bibitem{kapusta} 
  J.I.~Kapusta,
  {\it Quantum Chromodynamics at High Temperature,}
  Nucl.\ Phys.\ B {148} (1979) 461.
  %%CITATION = NUPHA,B148,461;%%


\end{thebibliography}
\end{document}